\documentclass{nature}

\newcommand{\arcsec}{''}

\usepackage{graphicx,hyperref,aas_macros,amssymb}

\begin{document}

\title{A planet-forming circumbinary disc in a polar configuration}

\author{Grant M. Kennedy$^1$, Luca Matr\`a$^2$, Stefano Facchini$^3$,
  Julien Milli$^4$, Olja Pani\'c$^5$, Daniel Price$^6$, David J.
  Wilner$^2$, Mark C. Wyatt$^7$, Ben M. Yelverton$^7$}

\maketitle

% List of institutions
\begin{affiliations}
\item Department of Physics and Centre for Exoplanets and Habitability, University of Warwick, Gibbet Hill Road, Coventry CV4 7AL, UK
\item Harvard-Smithsonian Center for Astrophysics, 60 Garden Street, Cambridge, MA 02138, USA
\item Max-Planck-Institut f\"ur Extraterrestrische Physik, Giessenbachstrasse 1, D-85748 Garching, Germany
\item European Southern Observatory (ESO), Alonso de C\'ordova 3107, Vitacura, Santiago, Chile
\item School of Physics \& Astronomy, University of Leeds, Woodhouse Lane, Leeds, LS2 9JT, UK
\item Monash Centre for Astrophysics (MoCA) and School of Physics and Astronomy, Monash University, Clayton, Vic 3800, Australia
\item Institute of Astronomy, University of Cambridge, Madingley Rd, Cambridge CB3 0HA, UK
\end{affiliations}

\noindent Do not read this, it is the original \textbf{submitted}
version, which is different to the accepted version.

\noindent The accepted version can be viewed for free at:
\href{https://rdcu.be/bgNSO}{https://rdcu.be/bgNSO}.

\clearpage

\begin{abstract}

  The circumbinary planets found to date by the transit method have
  orbits that are well-aligned with the binary orbital plane
  \cite{2011Sci...333.1602D,2012Natur.481..475W,2016ApJ...827...86K}. Misaligned
  circumbinary planets could exist and remain undiscovered because they
  are much harder to detect
  \cite{2014MNRAS.444.1873A,2014A&A...570A..91M}. The gas and dust rich
  discs in which planets form are initially present around nearly all
  young stars, but there has previously been no evidence of circumbinary
  discs that could form misaligned planets. Here we report the discovery
  of a strongly misaligned circumbinary disc that has properties similar
  to other planet-forming discs. Torques from the binary have forced
  this disc to a polar configuration that is stable
  \cite{2017ApJ...835L..28M}, providing the first evidence that
  misaligned circumbinary discs in this configuration exist and can
  survive long enough to form misaligned planets. The existence of
  planets on polar orbits could mean that that the occurrence rate of
  circumbinary planets is higher than for single stars.
  
  % Circumbinary planets orbit not one, but two stars, and nearly a dozen
  % have been discovered to date by the transit method
  % \cite{2011Sci...333.1602D,2012Natur.481..475W,2016ApJ...827...86K}. The
  % orbits of these planets are coplanar with the binaries they orbit, but
  % this characteristic arises because they are discovered in eclipsing
  % binary systems. Planets with non-coplanar orbits could exist, and not
  % have been discovered because they are much harder to detect
  % \cite{2014MNRAS.444.1873A,2014A&A...570A..91M}. The gas and dust rich
  % discs in which planets form are initially present around nearly all
  % young stars, but there has previously been no evidence of circumbinary
  % discs that could form misaligned planets. Here we report the discovery
  % of a strongly misaligned circumbinary disc that has properties similar
  % to other planet-forming discs. Torques from the binary have forced
  % this disc to a polar configuration that is stable
  % \cite{2017ApJ...835L..28M}, providing the first evidence that
  % misaligned circumbinary discs in this configuration exist and can
  % survive long enough to form misaligned planets. Although the precision
  % is limited by the small number of discoveries, coplanar circumbinary
  % planets are consistent with being as common as the equivalent planets
  % around single stars \cite{2014MNRAS.444.1873A}. Given the possibility
  % of planets on polar orbits, the occurrence rate of circumbinary
  % planets may therefore be higher than for single stars, and
  % circumbinary planet formation may proceed with greater efficiency than
  % around single stars.

\end{abstract}

\clearpage

To date, eleven circumbinary planets have been discovered by the transit
method in nine eclipsing binary systems that have semi-major axes of
between 0.08 and 0.23 astronomical units
\cite{2015MNRAS.453.3554M,2016ApJ...827...86K}. The strongest bias that
this discovery technique introduces is an almost total inability to
discover planets whose orbital planes are not closely aligned with that
of the binary \cite{2014MNRAS.444.1873A,2014A&A...570A..91M}. Despite
this bias, the occurrence rate of aligned circumbinary planets is
comparable to the frequency of equivalent planets around single stars,
albeit with considerable uncertainties because of the small numbers
involved \cite{2014MNRAS.444.1873A,2014A&A...570A..91M}. If unseen
misaligned planets also exist, circumbinary planet formation might be
more successful than around single stars.

While direct imaging surveys for circumbinary planets also exist, and
find consistent overall binary and single-star occurrence rates (albeit
with only 1-2 detections) \cite{2018arXiv180708687A}, they cannot
currently draw conclusions related to alignment because the planet
orbital periods are too long to derive orbits
\cite{2014ApJ...780L...4B}.

A quirk of the dynamics means that planet formation in misaligned
circumbinary discs could be successful. For small misalignments the
angular momentum vector of circumbinary orbits $\bf{L_{c}}$ precesses
about the binary angular momentum $\bf{L_b}$ (a `coplanar' family of
orbits), but for larger misalignments $\bf{L_{c}}$ precesses about a
vector in the binary's pericentre direction $\bf{\varpi_b}$ (a `polar'
family)\cite{2010MNRAS.401.1189F}. Because planet-forming discs are
gas-rich and hence dissipative, an initially misaligned circumbinary
disc will generally evolve to an end state that belongs to one of these
two families
\cite{2017ApJ...835L..28M,2018MNRAS.473..603Z,2018MNRAS.479.1297M}. While the
circumbinary debris disc 99~Herculis is thought to have a polar
configuration \cite{2012MNRAS.421.2264K}, this disc is four times larger
than Neptune's orbit, and with an age similar to the Solar system does
not provide evidence that misaligned circumbinary discs exist at the
epoch of planet formation.

Here, we report that the circumbinary disc in the young HD~98800 system
is strongly misaligned with the binary orbital plane. We infer that the
disc is in the polar configuration by simulating the disc dynamics. We
further show that despite this misalignment, the disc otherwise shows
physical properties similar to discs seen around young single
stars. Because a large fraction of single stars are known to host
planets, we may infer that single-star discs are in general successful
at forming planets, and therefore that similar processes are probably
ongoing in the circumbinary disc in the HD~98800 system.

\begin{center}
	\begin{figure*}
		\includegraphics[width=1\textwidth]{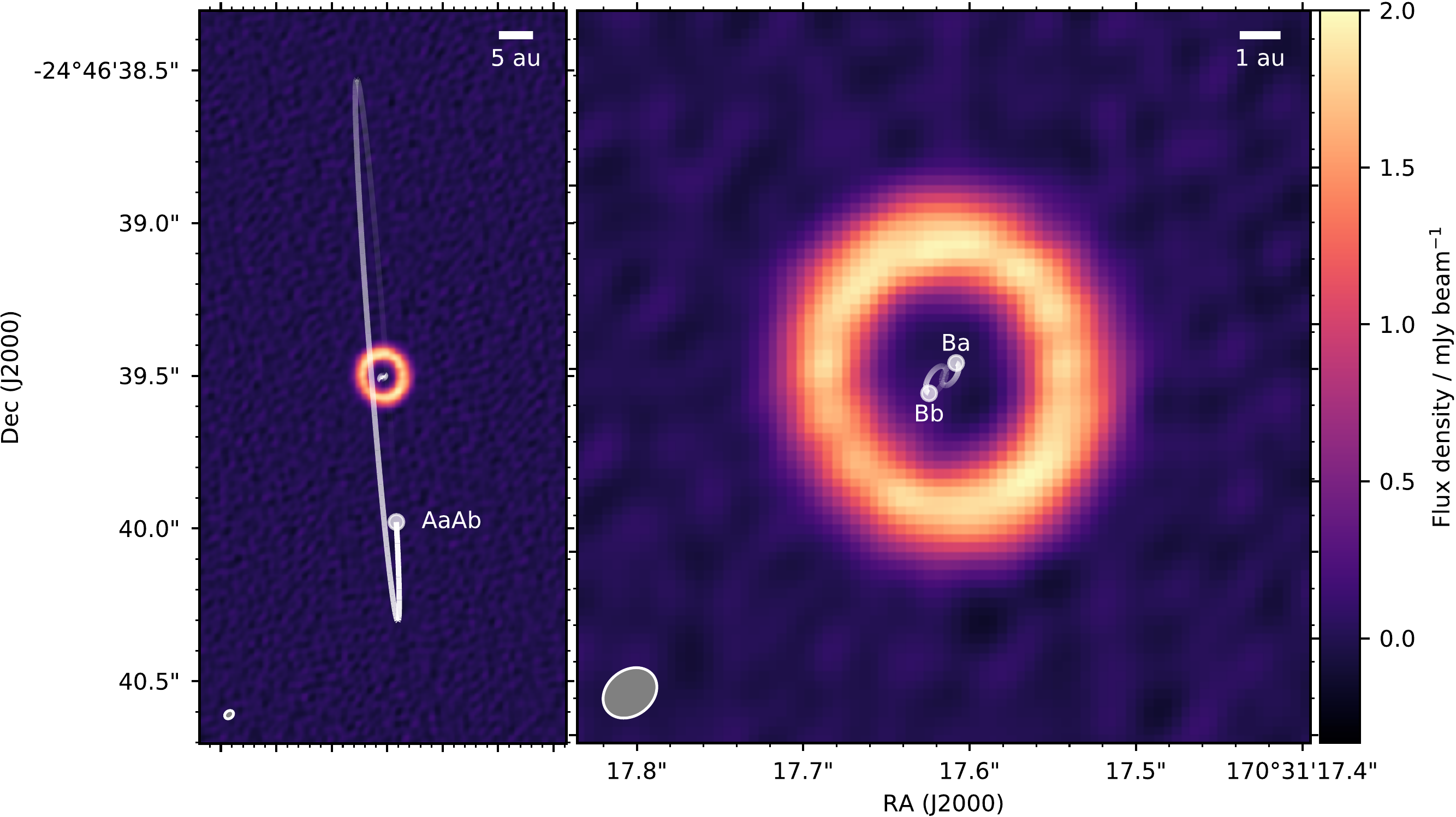}
		\caption{ALMA 1.3 millimetre continuum image of the
                  HD~98800 dust disc, showing a narrow dust ring 3.5~au
                  in radius that is 2~au wide. White semi-transparent
                  lines show the orbits of the inner binary (BaBb) and
                  the path of the outer binary (AaAb) with respect to
                  BaBb, with dots at the star locations at the time of
                  the ALMA observation. The resolution of these
                  `uniformly-weighted' images ($32 \times 25$
                  milli-arcseconds, or $1.4 \times 1.1$au) is given by
                  the ellipse in the lower left corner. The left panel
                  shows the entire system, and the right panel is zoomed
                  in on BaBb.}\label{fig:img}
	\end{figure*}
\end{center}

The HD~98800 system is a well known hierarchical quadruple star system
44.9~parsecs from Earth \cite{2007A&A...474..653V}, and a member of the
$\sim$10~Myr-old TW~Hydrae Association
\cite{1997Sci...277...67K,2006A&A...459..511B}. It consists of two pairs
of binaries (called `A' and `B', or equally `AaAb' and `BaBb') with
semi-major axes of about 1 astronomical unit (au), which themselves
orbit each other with a semi-major axis of 54~au. The binary BaBb is
well characterized, with an eccentricity of $0.785 \pm 0.005$, ascending
node of $289 \pm 1^\circ$ anti-clockwise from North, and inclination of
$67 \pm 3^\circ$ \cite{2005ApJ...635..442B}. Using the new data
presented here we derive a new orbit for AB (see Methods), which has an
eccentricity of $0.52 \pm 0.01$ and a period of $246 \pm 5$~years, with
an ascending node of $4.7 \pm 0.2^\circ$ and inclination of
$88.4 \pm 2^\circ$. These orbits as projected on the sky plane are shown
in Figure \ref{fig:img}. The AaAb orbit is less certain, but the details
do not affect our conclusions.

The Northern pair known as HD~98800BaBb has been known to host a bright
circumbinary disc since discovery in the 1980s
\cite{1988PASP..100.1509W,1992MNRAS.255P..31S,1993ApJ...406L..25Z}. The
disc is thought to be influenced by the stellar system
\cite{2007ApJ...670.1240A}, with the inner edge of the disc truncated by
the inner binary BaBb, and the outer edge externally truncated by A
\cite{2010ApJ...710..462A}. The orientation of the disc was initially
thought to be coplanar with the inner binary \cite{2010ApJ...710..462A},
but higher resolution observations suggest a different orientation
\cite{2018ApJ...865...77R} (see Methods for a comparison with our
results). Whether the disc harbours a significant mass of gas has been
unclear, meaning that it has been interpreted as both a gas-rich
`planet-forming' disc \cite{2007ApJ...664.1176F,2018ApJ...865...77R}, and a
gas-poor `debris' disc
\cite{2007ApJ...658..569W,2008MNRAS.390.1377V}. Detection of oxygen
towards the system \cite{2013A&A...555A..67R}, and molecular hydrogen
emission towards B \cite{2012ApJ...744..121Y}, suggests that this pair
is accreting from a gas-rich disc, favouring the former interpretation.

\begin{center}
	\begin{figure}
		\includegraphics[width=0.5\textwidth]{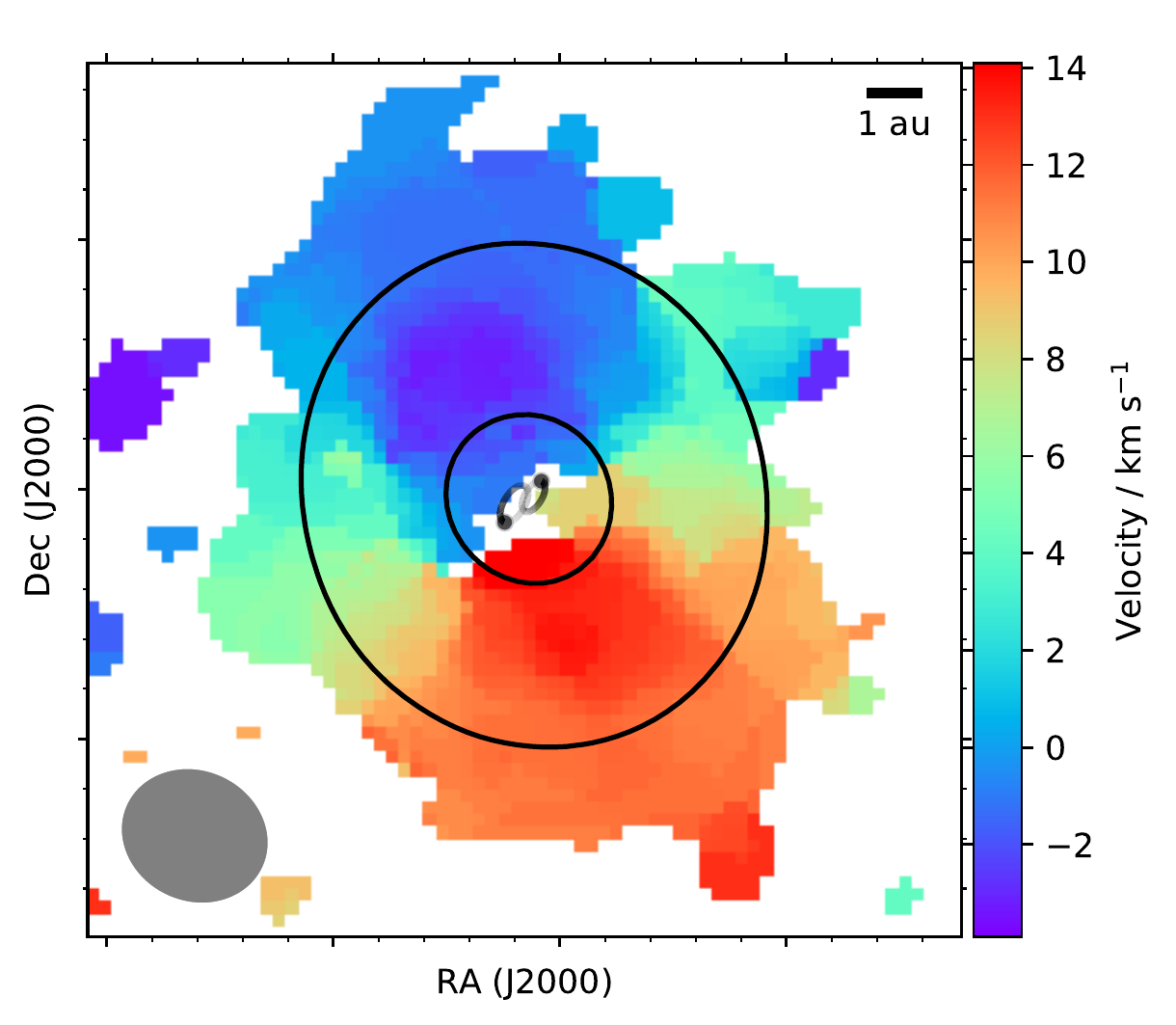}
		\caption{Intensity-weighted CO velocity (colours) and
                  dust (contours) images. The gas velocity structure is
                  consistent with that expected if the CO disc has the
                  same orientation as the dust. The CO is only shown
                  where it is detected at a signal-to-noise ratio
                  greater than three, and the contours are shown at 20
                  times the noise level in the continuum image. The
                  resolution of these `naturally-weighted' images
                  ($61 \times 54$~milli-arcseconds, or
                  $2.7 \times 2.4$~au) is given by the ellipse in the
                  lower left corner.}\label{fig:mom1}
	\end{figure}
\end{center}

To ascertain the disc orientation, size, structure, and evolutionary
status we observed the HD~98800 system with the Atacama Large
Millimeter/sub-millimeter Array (ALMA, see Methods). Data were taken at
230GHz (1.3~mm), to image dust continuum emission and the carbon
monoxide (CO) J=2-1 rotational transition, and both are strongly
detected (Figures \ref{fig:img} and \ref{fig:mom1}). By modelling these
data as a disc that lies between inner and outer radii with a power-law
surface brightness prescription, we find that the inner edges of the
dust and CO are at $2.5 \pm 0.02$ and $1.6 \pm 0.3$ au respectively,
while the outer edges are at $4.6 \pm 0.01$ and $6.4 \pm 0.5$ au (see
Methods). These models show that the dust and CO components are
consistent with having the same orientation; the disc has a position
angle of $16 \pm 1^\circ$ (measured anti-clockwise from North) and is
inclined by either 26 or 154$^\circ$ ($\pm$1$^\circ$) from the sky
plane. While the Doppler shifts seen in CO show that the North side of
the disc is rotating towards us (Fig. \ref{fig:mom1}), thus constraining
the ascending node to be North of the star, the inclination remains
ambiguous because the disc could be rotating either clockwise or
anti-clockwise as projected on the sky. That is, these observations do
not distinguish whether the East or West side of the disc is closer to
Earth.

\begin{center}
	\begin{figure}
		\includegraphics[width=0.5\textwidth]{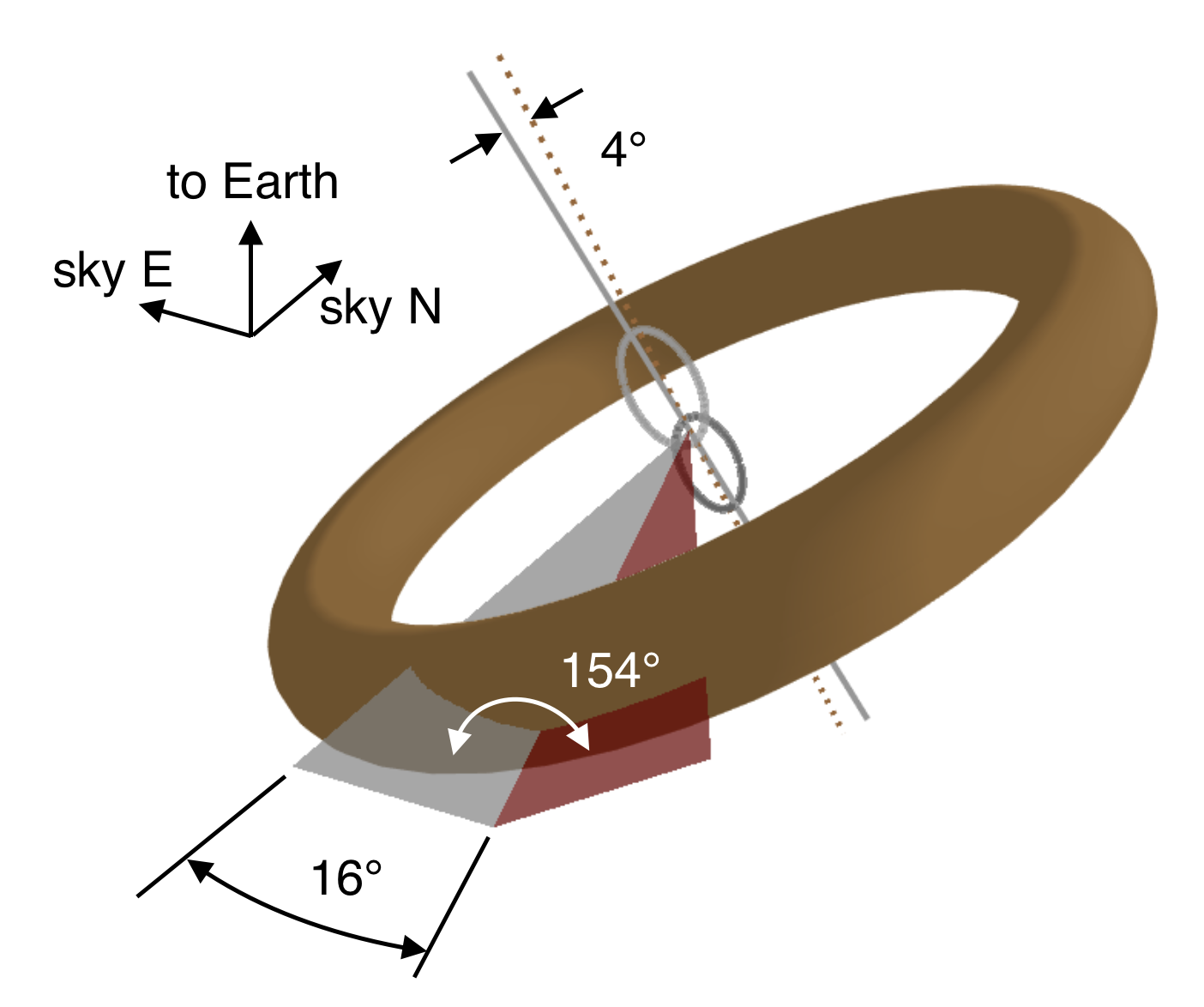}
		\caption{Three-dimensional sketch of the polar
                  configuration of the BaBb binary (grey ellipses) and
                  surrounding disc (torus). The axes are indicated by
                  the three labelled arrows. The two triangular planes
                  show the plane of the sky (grey) and of the disc
                  (red), which intersect at an angle of 16$^\circ$ from
                  South (i.e. the disc position angle). The angle
                  between these planes is the disc inclination of
                  154$^\circ$. Grey lines show the angular momentum
                  vector of the disc $\bf{L}_c$ (dotted), and the binary
                  pericentre direction $\bf{\omega}_b$ (solid), both of
                  which point towards away from Earth. These vectors are
                  aligned within 4$^\circ$, meaning that for this
                  orientation the disc plane is almost perfectly
                  perpendicular to both the binary orbital plane and the
                  pericentre direction.}\label{fig:3d}
	\end{figure}
\end{center}

Of the two possible disc orientations, the 154$^\circ$ case is only four
degrees away from the polar configuration (i.e. is perpendicular to both
the BaBb orbital plane and the BaBb pericentre direction), while the
26$^\circ$ case is inclined 48$^\circ$ from the BaBb binary plane (which
we refer to as the `moderately' misaligned case). Given the small chance
that a randomly chosen orientation should appear to be in the polar
configuration that it is expected based on the dynamics and models
discussed above, this configuration is by far the most likely
interpretation, and a sketch is shown in Figure \ref{fig:3d}.

To further test this hypothesis, we simulated the response of the disc
to perturbations from the stellar orbits using both gas-free `$n$-body',
and fluid-based smoothed particle hydrodynamics, simulations (see
Methods). The main conclusion from the $n$-body simulations is that test
particles placed on circular orbits at the observed 2.5-4.6~au radial
location of the dust are generally ejected within less than a million
years, regardless of the disc orientation. Independent of the disc
orientation we therefore conclude that the dust observed with ALMA is
embedded within a more massive gas disc, which acts to stabilize the
disc against both interior and exterior stellar perturbations. Our
estimates of the gas and dust mass in the disc are consistent with this
picture, but do not confirm it because they have large systematic
uncertainties (see Methods). The $n$-body dynamics, in concert with the
detection of oxygen and hydrogen gas, therefore suggest that
HD~988000BaBb almost certainly harbours a long-lived gas-rich
protoplanetary disc.

Our fluid based simulations find that torques from the inner binary
re-orient a gas disc in the moderately misaligned configuration to the
polar configuration in several hundred years. In contrast, the polar
configuration is stable and remains in the observed state for at least
500 BaBb orbits (430 years). We therefore conclude that the disc is in
the polar configuration, as the moderately misaligned configuration is
reoriented on a timescale that is very short relative to the ten
million-year old age of the system. The relatively fast realignment time
also suggests that the disc has been in this configuration for most of
its lifetime. Given that most stars disperse their discs well within ten
million years \cite{2001ApJ...553L.153H}, the HD~98800BaBb disc has had
ample opportunity to form planets on polar orbits.

While we cannot currently detect planets within the HD~98800 disc (the
high optical depth and small angular scale hamper imaging, and radial
velocity precision is limited for young binaries), the probable high
dust and CO optical depths are characteristics shared by all
protoplanetary disks. The physical conditions, which successfully lead
to planet formation around single stars, are therefore likely to be
similar for HD~98800. This system may therefore show the same key
signature of the first stages of planet formation observed in discs
around other stars; gas is commonly seen to extend farther from the star
than dust, which is thought to be caused by inward radial drift of dust
particles that have grown beyond microns in size, and/or the in-situ
growth of dust to sizes large enough that their mm-wave emission is
fainter at larger distances
\cite{2012ApJ...744..162A,2017A&A...605A..16F}. Evidence of dust growth
in the outer disc may be present in our ALMA observations, as our models
find that the gas is more extended than the bulk of the dust (see also
Fig. \ref{fig:mom1}). While the derived edge locations reflect the
sensitivity of the observations to some degree, these models account for
all of the CO surface brightness, and for 99\% of the dust. The dust
surface brightness is a factor of a hundred lower at 5.5au compared to
3.5au, which is much larger than the factor of two expected if the dust
emission scaled with the brightness of the gas emission over this
distance. Our models therefore show probable evidence of the effects of
grain growth on radial disc structure, and therefore the same evidence
of the first steps towards planet formation seen in equivalent discs
around single stars.

While the formation of circumbinary planets on polar orbits therefore
seems generally possible, the long-term prognosis for any planets that
form around HD~98800 is mixed. The gas currently protects solid bodies
orbiting within the disc from external perturbations from A, but as
shown by our gas-free simulations this protection will largely disappear
when the gas is dispersed. To survive, any planets that formed in this
specific system must either reside just outside 4~au or near 6~au, and
even these locations do not provide a sure refuge. More likely, any
planets are destined to be consumed upon colliding with one of the
systems' stars, or to float through space after ejection from the
system.

How did this system form? Simulations suggest that such outcomes may be
a natural result of the chaotic nature of star formation
\cite{2018MNRAS.475.5618B}. One possibility is that two initially
separate binary systems became gravitationally bound, and misaligned
HD~98800BaBb's disc in the process (and perhaps destroyed any disc
around A). However, simulations show that formation from molecular cloud
material with a range of angular momenta can result in misaligned discs
in isolated systems \cite{2018MNRAS.475.5618B}, and indeed such systems
are observed \cite{2016ApJ...830L..16B}, so it does not appear that the
presence of an exterior companion is a necessary condition for forming
misaligned circumbinary discs, or that formed planets will necessarily
be ejected later by perturbations from such companions. The IRS~43
system, with a circumbinary disk misaligned by about 60$^\circ$
\cite{2016ApJ...830L..16B}, is a possible example of what systems such
as HD~98800 could look like earlier in their evolution before binary
torques re-orient the disk (albeit on a scale of tens of au, rather than
a few au).

Our results imply that circumbinary planets may be more common than
currently thought.  Assuming discs that are initially randomly oriented
relative to the binary orbital plane, that are not massive enough to
affect the binary orbit, and that binary eccentricities are uniformly
distributed up to $e=0.8$ \cite{2010ApJS..190....1R}, the fraction of
discs that should evolve to a polar configuration is about 45\% (see
Methods). The most eccentric binaries are the most likely to have polar
disc configurations and it is therefore not surprising that the known
transiting circumbinary planets are all in systems with $e \le 0.52$,
with 8/9 having $e<0.22$
\cite{2015MNRAS.453.3554M,2016ApJ...827...86K}. While our population
estimate is simple, it shows that polar discs and planets may be a
common outcome of circumbinary disc formation.

\begin{methods}

\subsection{Observations and Data Processing}\label{sup:data}

% First: X7bf8, Second: X311b

% Continuum rms in uJy/beam:
%X7bf8, uniform: 79
% X311b, uniform: 60
% all, briggs: 14
% all, natural: 13
% all, uniform: 43
% spw04, natural: 20

HD~98800 was observed by ALMA in Band 6 (1.3~mm) in two observing blocks
on 2017 November 15 and 19, 49 antennas were used in the first block and
45 in the second. The shortest and longest baselines were 92~m and
11.8~km. The correlator configuration used three broad continuum windows
with 2GHz bandwidth, and one centred on the CO~J=2-1 line with a
spectral resolution of 488kHz (0.73km~s$^{-1}$ velocity
resolution). Each block comprised observations of HD~98800, interspersed
with observations of phase calibrator J1104-2431. J1127-1857 was used as
the bandpass and flux calibrator. The time on-source was 34 minutes for
each block. The raw data were calibrated using the observatory pipeline
with CASA v5.1.

The data were further processed using one cycle of phase
self-calibration (self-cal) to improve the signal to noise ratio
(s/n). All spectral windows were combined, and the averaging time
(`solint') was chosen to be relatively long (20-40min) to avoid flagging
antennas due to low s/n, thereby retaining maximal spatial resolution.
The rms variation measured in clean images before self-cal with Briggs
weighting (robust=0.5) was 27 and 32~$\mu$Jy~beam$^{-1}$ for the first
and second observations. After self-cal these were 20 and
18~$\mu$Jy~beam$^{-1}$. Attempts at further self-cal did not improve the
s/n.

Following calibration the two observations were concatenated into a
single set of visibilities for imaging and modelling. This combination
was verified to be reasonable by modelling the continuum of each
observation separately (as described below), which found that the sky
offsets of the disc were consistent to within 0.0002~arcsec
(0.01au). The difference in integrated flux densities was consistent to
within 1\%. In the final Briggs-weighted clean image the beam size is
$46 \times 42$mas, the rms is 14~$\mu$Jy~beam$^{-1}$, and the peak s/n
is 280. In a naturally weighted image the beam size is
$64 \times 56$mas, the rms is 13~$\mu$Jy~beam$^{-1}$, and the peak s/n
is 450. In a uniformly weighted image the beam size is
$32 \times 25$mas, the rms is 43~$\mu$Jy~beam$^{-1}$, and the peak s/n
is 46.

For continuum modelling a single spectral window from the combined
observations was used (the first, centred at 1.311 mm) with the 2GHz
bandwidth averaged into four channels. While more data could have been
used, the continuum s/n is easily sufficient to obtain stringent
modelling constraints with one spectral window. Visibility data were
time averaged into 20s chunks, and then exported to a text file and
modelled as outlined below.  The weights associated with each visibility
were divided by a re-weighting factor such that a null model produces a
$\chi^2$ value of 1, based on the expectation that each individual
visibility measurement has negligible s/n
\cite{2011A&A...529A.105G}. The re-weighting factor was 5.8 (which is
decreased to 2.6 if the CASA \texttt{statwt} task is run first).

For gas modelling the window centred on the CO~J=2-1 line was used. The
continuum was subtracted using the CASA \texttt{uvcontsub} task, the two
observations merged using the \texttt{mstransform} task, and 40 channels
near the CO line extracted to a series of text files containing the
visibilities at each frequency (or equivalently, velocity). A
re-weighting factor was again used, which was within a few percent of
one. A naturally-weighted clean CO cube has a beam size of
$61 \times 54$~mas, an rms of 0.8~mJy~beam$^{-1}$ in each
0.73~km~s$^{-1}$ channel, and typical peak s/n of 4 to 6 depending on
the channel.

HD~98800 has been observed at millimetre wavelengths many times in the
past with single-dish telescopes. These show a considerable degree of
scatter, with ref \cite{1995Ap&SS.224..389W} measuring
$30.7 \pm 8.2$~mJy and $54.4 \pm 3.61$~mJy at 1.3mm with the CSO and
IRAM respectively. Our modelling below yields a total disc flux of
$47.4 \pm 0.4$~mJy. Including an absolute calibration uncertainty of
10\% yields a final flux measurement of $47 \pm 5$~mJy. Our value is not
significantly below the previous single dish measurements, so it is
unlikely that we have resolved out significant flux.

The phase centre of the observations is not perfectly centred on either
AaAb, BaBb, nor the system photocentre, no doubt caused by some
uncertainty in the actual position of the system components and their
relative motions as derived by Hipparcos \cite{2007A&A...474..653V}. As
another output of the modelling, assuming that the BaBb barycentre is at
the centre of the disc, at the time of the observation (2017.874) we
find that BaBb is centred at $\alpha = 11$~22~05.17437,
$\delta=-24$~46~39.5043 (170.52155986$^\circ$,
-24.77764009$^\circ$). The positional uncertainty from the modelling is
$\pm 0.0001$~arcsec, but the true uncertainty is limited by ALMA's
pointing accuracy, which the Technical Handbook suggests is about
0.03~arcsec.

\subsection{Visibility Modelling}\label{sup:mod}

Modelling of the continuum and CO data was done in the visibility plane
using an optically thin line of sight integration code. While the dust
is likely optically thick, the use of a radiative transfer code would
make little difference here because the disc is close to face-on, and we
are therefore simply modelling the surface brightness of the disc as a
function of radius. A function specifies the three-dimensional disc
density in spherical polar coordinates, which is mapped into a 3d
cartesian volume using two or three rotations (the position angle
$\Omega$, the inclination $i$, and the argument of pericentre $\omega$
where necessary). Two axes of this cube represent the sky plane and the
third the line of sight, and the final continuum image of a given model
is created by summing the cube along the line of sight axis. Velocity
cubes of the model are created by first computing the radial velocity at
each location in the cube. Layers in the velocity cube are again the sum
along the line of sight axis, but only including pixels from the cube
that are within the velocity range for a given layer.

The range of models that are consistent with the data are found using
the Markov-Chain Monte-Carlo (MCMC) package \texttt{emcee}
\cite{2013PASP..125..306F}. The log likelihood $-\chi^2/2$ of each
model is computed given the visibility data using GALARIO
\cite{2018MNRAS.476.4527T}. GALARIO also computes the pixel size and
image extent necessary for sufficient $u,v$ resolution when the images
are transformed into visibilities, which are 4.6~milli arcseconds per
pixel and 2048~pixels.

The strongest signal in the continuum image in Figure \ref{fig:img} is a
narrow ring of dust emission, so we model this ring to derive
constraints on structure and reveal any lower-level emission. For both
continuum and CO we use a simple power-law density model, where the dust
or gas lies between two limiting radii $r_{in}$ and $r_{out}$ and where
the volume density is a power law function of radius $r^\alpha$. The
density is specified in a given pixel in the cube, and the vertical
scale height is a fixed fraction of the radial distance, so the surface
density is $\propto r^{\alpha+1}$. The vertical density structure is
Gaussian with the scale height fixed to $0.05r$ for all models, as for a
nearly face-on disc this parameter is poorly constrained. For the
continuum model we found that the fit was significantly improved if the
disc inner edge has a small eccentricity $e_{in}$, which was implemented
by varying $r_{in}$ as a function of azimuth. Further parameters are the
sky offsets $x_0$ and $y_0$, the disc position angle East of North
$\Omega$ and inclination $i$, and the argument of pericentre
$\omega$. The images are scaled by the total flux $F$ (in the 2d image
for continuum, or the 3d cube for CO). There are therefore eight
parameters for the continuum model ($x_0$, $y_0$, $\Omega$, $\omega$,
$i$, $F$, $r_{in}$, $r_{out}$, $\alpha$, and $e_{in}$).  The density is
multiplied by an emission function that mimics the Rayleigh-Jeans tail
of a blackbody (i.e.  $\propto 1/\sqrt{r}$). The radial dependence of
this function is largely arbitrary as it is degenerate with $\alpha$,
and while it is a reasonable approximation for the dust seen in the
continuum, the temperature dependence for CO may be different. For the
CO model the eccentric inner edge parameters $\omega$ and $e_{in}$ are
not necessary as the s/n is much lower, but an additional parameter, the
systemic velocity of HD~98800~BaBb, $v_{sys}$, is needed. We assume the
CO orbits a single point mass with the combined mass of HD~98800BaBb of
1.28$M_\odot$ \cite{2005ApJ...635..442B}.

While it is likely that the CO and dust are optically thick we found
that our models were sufficient to reproduce the data. We tried
modelling the data with a simple optical depth prescription where the
observed surface brightness in the model sky images was attenuated by a
further free parameter $\tau_{SB}$ via $1-e^{-SB/\tau_{SB}}$
(i.e. $\tau_{SB}$ is the surface brightness at which the emission
becomes optically thick), but we found that these models were consistent
with $\tau_{SB}=0$ (and the other parameters unchanged) and therefore
that this additional complication is not necessary to obtain constraints
on the disc extent and spatial orientation.

\begin{center}
	\begin{figure*}
		\includegraphics[width=1\textwidth]{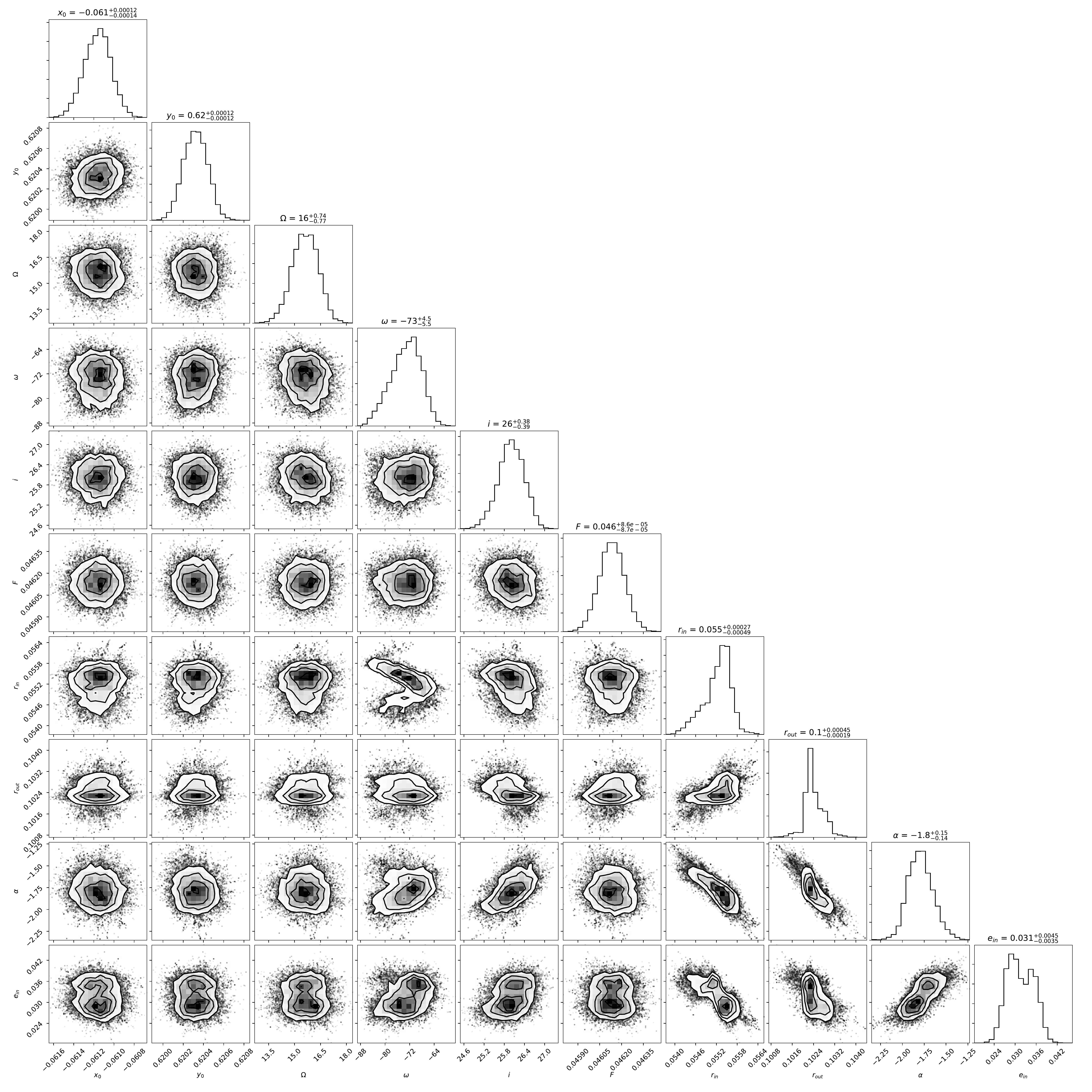}
		\caption{Posterior distributions for the continuum
                  model. The units of $x_0$, $y_0$, $r_{in}$ and
                  $r_{out}$ are seconds of arc, $\Omega$ and $i$ are
                  degrees, $F$ is Jansky, and $\alpha$ and $e_{in}$ are
                  dimensionless. The distance to HD~98800 is 44.9pc
                  (i.e. $r_{in}=2.5$au).}\label{fig:cont_mod}
	\end{figure*}
\end{center}

\begin{center}
	\begin{figure*}
		\includegraphics[width=0.48\textwidth]{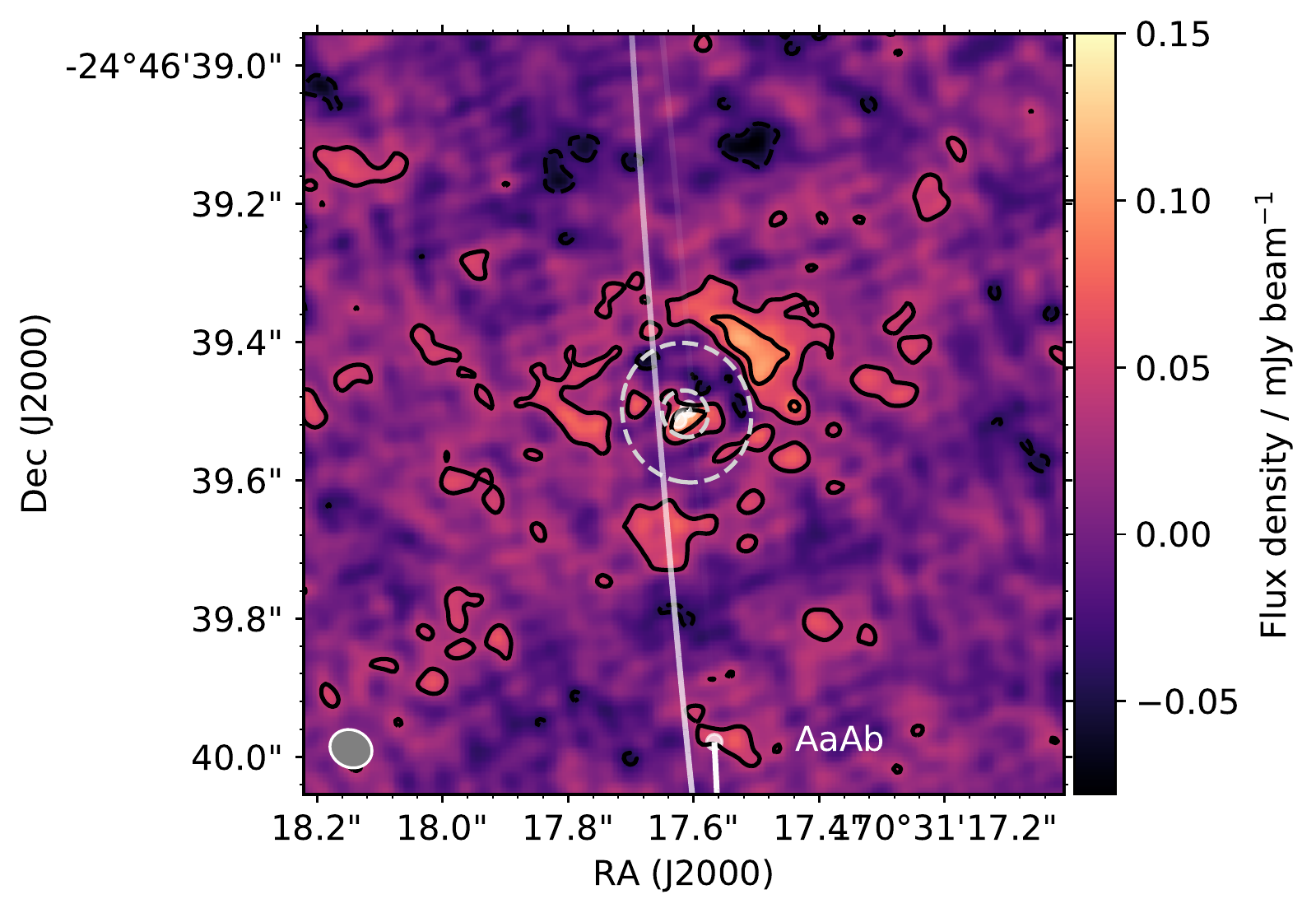}
		\includegraphics[width=0.48\textwidth]{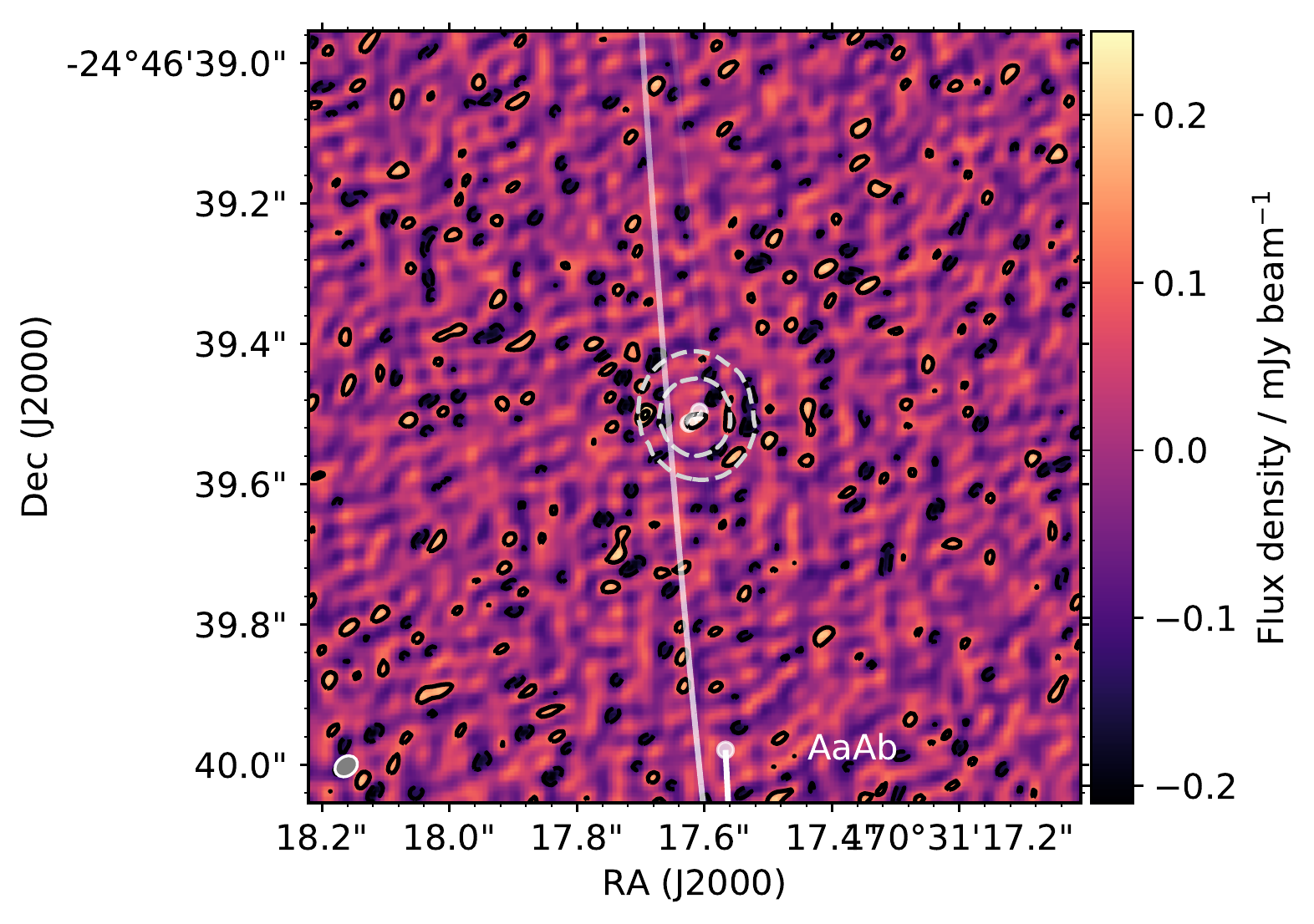}
		\caption{Residuals after subtracting the best fitting
                  continuum model. The left panel shows the image with
                  natural weights with solid contours at $\pm$2 and
                  4$\sigma$ (with $\sigma=$20~$\mu$Jy~beam$^{-1}$), and
                  dashed contours at 20$\sigma$ from the original
                  image. As in other images the stellar orbits are also
                  shown. The right panel is the same as the left panel,
                  but for uniform weights, and with
                  $\sigma=$60~$\mu$Jy~beam$^{-1}$.}\label{fig:cont_resid}
	\end{figure*}
\end{center}

\emph{Continuum modelling results:} The best fitting continuum model was
found using 64 `walkers' (parallel MCMC chains) run for 1000 steps
(having discarded a prior `burn-in' run of 1000 steps). The posterior
distributions and their correlations are shown in Figure
\ref{fig:cont_mod}. The y offset $y_0$ is relatively large because the
observation phase centre is offset from BaBb as described above. The
main parameters of interest here are the disc orientation $\Omega$ and
$i$, and the dust extent from $r_{in}$ to $r_{out}$, which are discussed
in the main text.

The flux density from the ring of 46~mJy can be converted to a dust
mass, if the emission is assumed to be optically thin. Using equation 5
of ref \cite{2017MNRAS.470.3606H} (which assumes an opacity of
1.7~cm$^2$~g$^{-1}$) and assuming a dust temperature of 160K derived
from photometry over a range of wavelengths \cite{1993ApJ...406L..25Z},
the result is $0.33M_\oplus$. Whether the dust ring is actually
optically thick, and therefore whether the dust mass is underestimated,
is uncertain; the flux expected from a ring of uniform surface
brightness extending from 2.5 to 4.5au at a constant temperature of 160K
is 130~mJy. This estimate does not mean the dust is optically thin
however, as the surface density profile is found to be decreasing with
radius. The bulk of the dust emission may therefore be concentrated in
an optically thick region that is narrower than the best-fit 2au width,
an issue that can be resolved with higher spatial resolution
observations.

Initial attempts to model the continuum with axisymmetric models left
asymmetric residuals suggestive of an offset near the inner disc edge,
which motivated the use of an eccentric inner disc edge. While this
parameterisation may not be representative of the true structure (which
could for example be similar to the simulations shown in Figure
\ref{fig:sph}), our finding that the best-fit $e_{in}$ is significantly
greater than zero shows that the disc is not axisymmetric, and that this
asymmetry lies near the inner disc edge. The position angle of the
pericentre is given by the parameter $\omega$; as it is measured from
the ascending node the best-fit value of $-73^\circ$ means that the disc
is closest to the binary towards the North-West (i.e. a position angle
of approximately $-72+16=-56^\circ$). It is likely that the preference
for a non-zero inner edge eccentricity results from perturbations from
the inner binary, and the structure may be characterized in more detail
with higher resolution imaging.

Residual images are shown in Figure \ref{fig:cont_resid}, which shows
that our model reproduces nearly all of the observed continuum
structure. The peak s/n in the original naturally weighted continuum
image for the same spectral window is 300, meaning that the residuals
are at most only 1\% of the peak. Some emission remains beyond the
bright ring at 2-4$\sigma$, suggestive of low-level dust emission that
might be recovered more strongly in lower resolution images. Some
residual emission is also seen interior to the ring, which might arise
from dust entrained in gas that is flowing towards and accreting onto
the inner binary, and be related to our finding that the inner disc edge
is asymmetric. Accreting material may provide an explanation for the
facts that HD~98800 was reported to be photometrically variable by
Hipparcos \cite{1998ApJ...498..385S}, and that HD~98800B suffers
significantly more dust extinction than A
\cite{1998ApJ...498..385S,1999AstL...25..669T,2005ApJ...635..442B,2007ApJ...670.1240A}.

% Inspection of public SuperWASP data from the NASA Exoplanet Archive does
% not show obvious dimming events, but a clear 14.2~day periodicity with a
% $\sim$4\% amplitude is visible. The Hipparcos epoch photometry shows a
% local peak at a period of 14.7~days and amplitude of 0.025~mag (stronger
% periods are present in the data, but some are presumably related to the
% sparse data sampling of 123 observations over about three years). These
% results therefore agree with the 14.8~day period found by
% \cite{1995AJ....110.2926H}. Removing the periodic signal from the
% Hipparcos data lowers the standard deviation (where one discrepant point
% has been excluded) from 0.027~mag to 0.023~mag, but HD~98800 remains
% more variable (according to the \texttt{Hpscat} column) than 85\% of
% stars of a similar magnitude.

\begin{center}
	\begin{figure*}
		\includegraphics[width=0.9\textwidth]{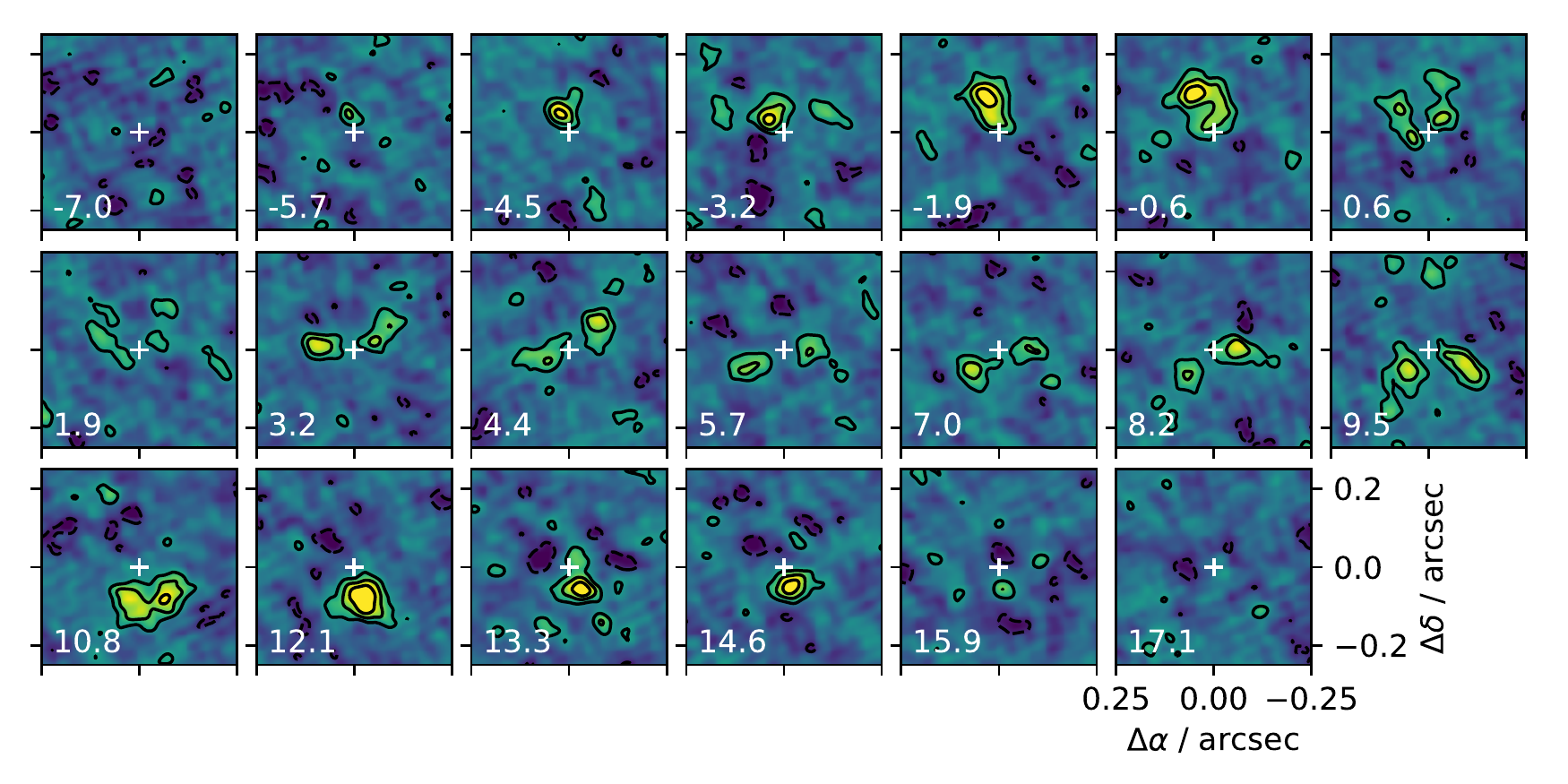}
		\includegraphics[width=0.9\textwidth]{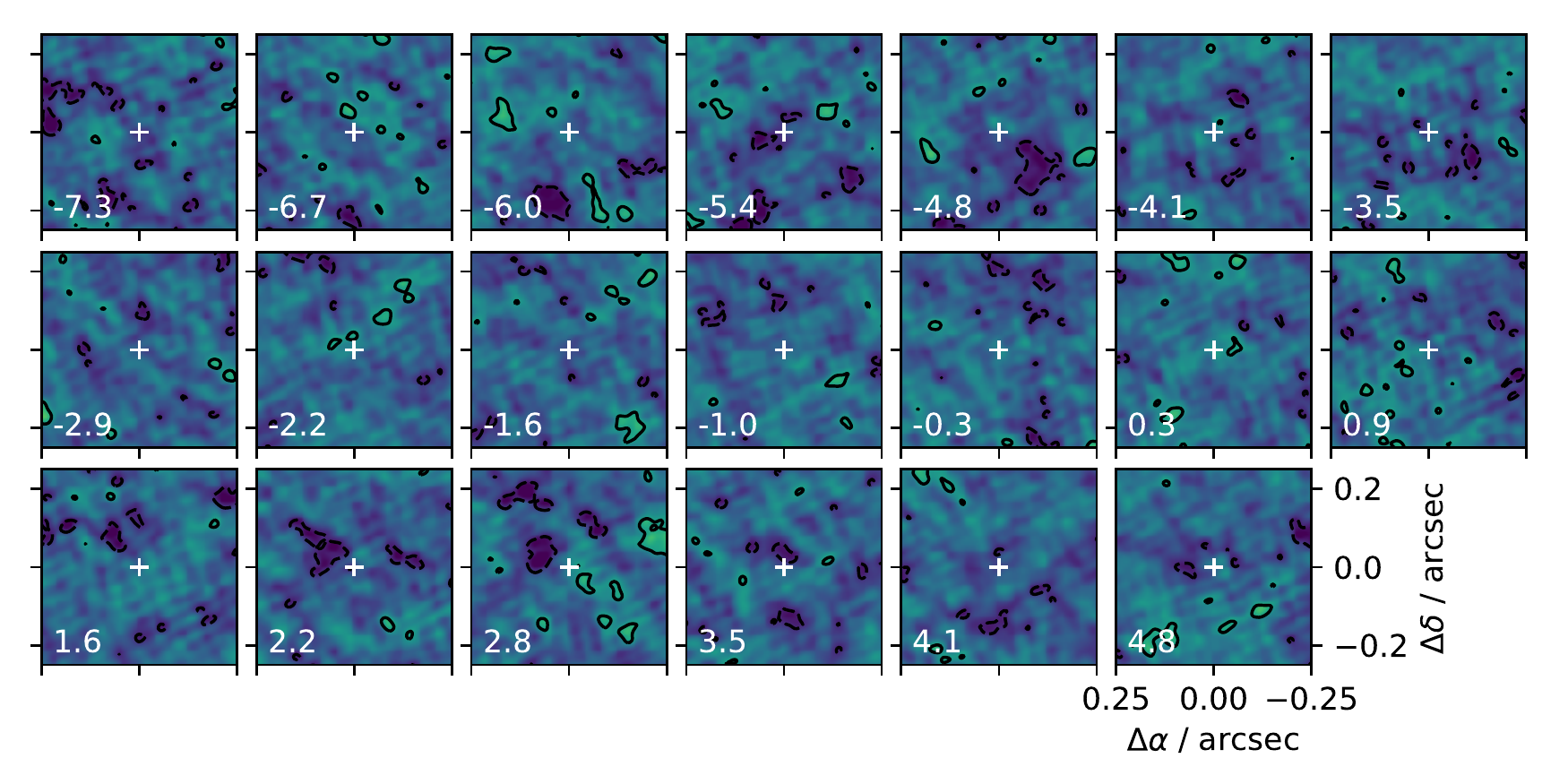}
		\caption{CO channel maps and residuals. The upper panel
                  shows the original cube with natural weights, where
                  each sub-panel is labelled with the velocity on which
                  the bin is centred (in km~s$^{-1}$). The lower panel
                  is the same as the upper panel, but the best-fitting
                  model has been subtracted. Contours are shown at -2,
                  2, 4, and 6$\sigma$, where
                  $\sigma=0.8$~mJy~beam$^{-1}$ and the beam size is
                  $0.06 \times 0.05$\arcsec.}\label{fig:co_resid}
	\end{figure*}
\end{center}

\begin{center}
	\begin{figure*}
		\includegraphics[width=1\textwidth]{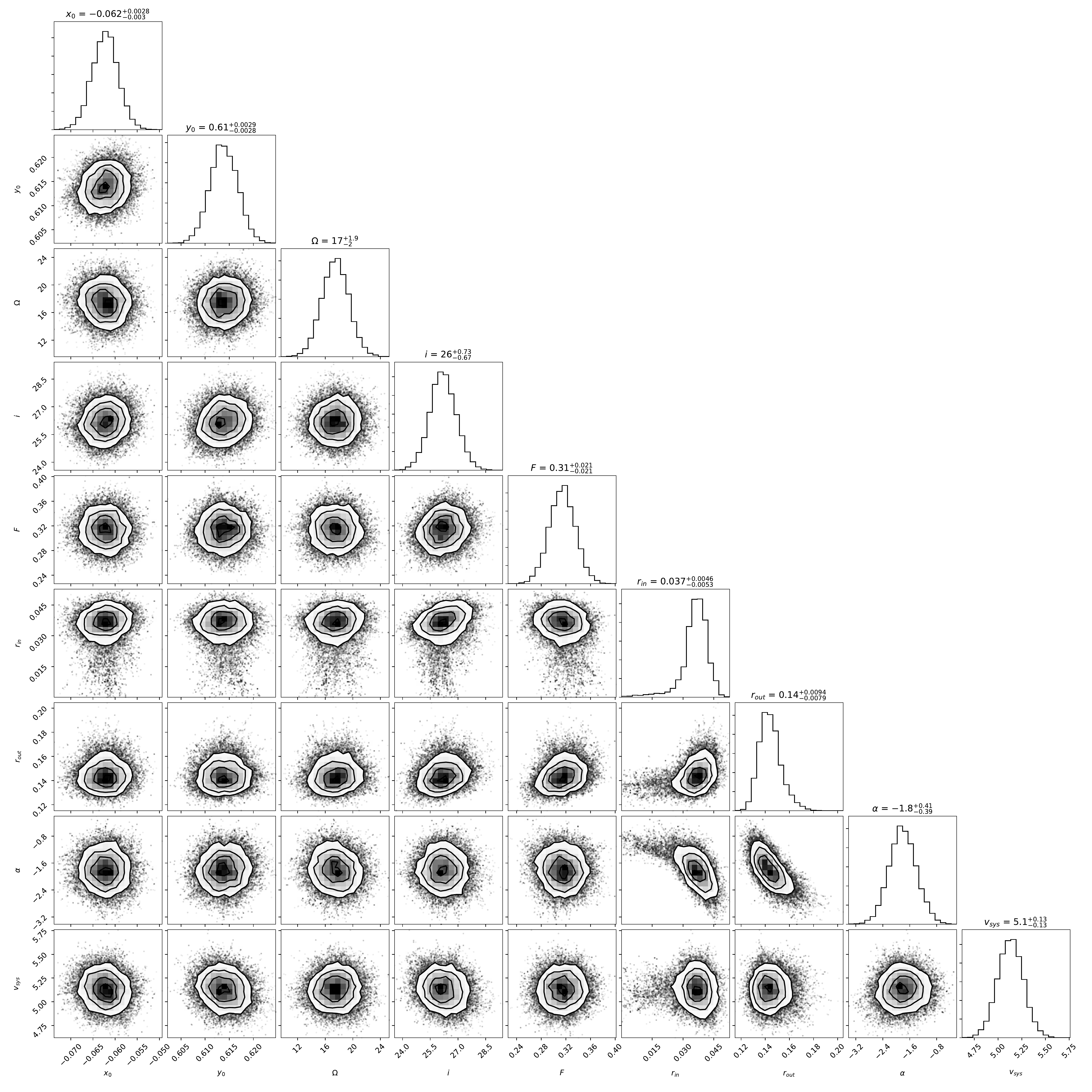}
		\caption{Posterior distributions for the kinematic CO
                  model. The units of $x_0$, $y_0$, $r_{in}$ and
                  $r_{out}$ are seconds of arc, $\Omega$ and $i$ are
                  degrees, $F$ is Jansky~km~s$^{-1}$, $\alpha$ is
                  dimensionless, and $v_{sys}$ is
                  km~s$^{-1}$.}\label{fig:co_mod}
	\end{figure*}
\end{center}

\emph{CO modelling results:} Channel maps near the systemic velocity of
HD~98800BaBb are shown in the left panel of Figure
\ref{fig:co_resid}. The best fitting CO velocity model was found using
64 walkers run for 1000 steps, where 1000 prior steps were discarded as
a burn-in. The posterior distributions and their correlations are shown
in Figure \ref{fig:co_mod}. The residuals after subtraction of the
best-fitting model are shown in the right panel of Figure
\ref{fig:co_resid}. The CO disc orientation is consistent with the
continuum model results shown in Figure \ref{fig:cont_mod}, but the CO
disc covers a greater radial extent, from 1.6 to 6.4au compared to 2.5
to 4.6au.

The peak level of channel emission in Figure \ref{fig:co_resid} is
approximately 5~mJy~beam$^{-1}$, which we find is consistent with that
expected from optically thick CO at an excitation temperature of 70K. A
lower excitation temperature would yield a lower flux than observed,
while a higher excitation temperature would yield a higher flux, or
could originate from optically thin CO.

We can estimate a minimum CO mass by assuming that the CO emission is
marginally optically thick ($\tau=1$) with an excitation temperature of
70K (i.e. this mass estimate only applies if $\tau \gtrsim 1$). The
best-fit total CO flux of 0.31~Jy~km~s$^{-1}$ is then equivalent to a
mass of $2.8 \times 10^{-5} M_\oplus$. If we assume that the disc is
primordial and has a standard H$_2$/$^{12}$CO ratio of 10$^4$, then the
total gas mass is roughly 0.28$M_\oplus$, similar to the dust mass.

In terms of the dust to gas mass ratio, the empirical evidence for which
dominates remains uncertain because i) even if both are optically thick,
the dust optical depth could be much greater than the CO optical depth
and thus the dust mass greater than the gas mass (or vice versa), ii)
the CO excitation temperature may be higher than our 70K estimate, in
which case the gas would be optically thin and the mass lower than our
minimal estimate, and iii) the ratio of molecular hydrogen to CO could
be different to our assumed value, in which cas the total gas mass could
be higher or lower than our estimate, even if the CO mass is
correct. Further observations that target other $^{12}$CO transitions,
and optically thinner lines such as $^{13}$CO and C$^{18}$O, are needed
to provide further information on the disc optical depth and gas
temperature.

\subsection{Comparison with VLA}\label{sup:vla}

Ribas et al. (ref \cite{2018ApJ...865...77R}) imaged the HD~98800B disk with
the Karl G. Jansky Vary Large Array (VLA), and derived constraints on
the continuum disk size and orientation (they did not detect CO). They
report a disk inclination of 40-45$^\circ$ at a position angle of
0-10$^\circ$, but do not quote formal confidence intervals, and do not
discuss how this orientation compares to the binary orbital
planes.\footnote{The original version posted to arXiv had
  PA=90-100$^\circ$, but we verified with the authors that this is a
  typo.} We modelled the VLA image using the model described above
(though with $\alpha$ fixed in the range -1 to 0, as the s/n is much
lower), and find a disk extent consistent with that derived from ALMA,
and an inclination of $35 \pm 7^\circ$ and position angle
$9 \pm 10^\circ$, also consistent with our results.

\subsection{Orbits}\label{sup:orb}

Ref \cite{1995ApJ...452..870T} find that the systemic velocity of AaAb
(12.75~km~s$^{-1}$) was more positive than BaBb (5.73~km~s$^{-1}$) in
the early 1990s, when A and B were near maximum elongation. Thus, at
that time BaBb was moving towards Earth relative to the system centre of
mass. The ascending node of the AB orbit (measured East of North) as
reported in previous literature is therefore incorrect and should be
near 4.2$^\circ$, not 184.2$^\circ$
\cite{1995ApJ...452..870T,2014AJ....147..123T}. Thus, as indicated in
the figures, AaAb will go behind the disc in 2026. AaAb being a binary,
time-series photometry may reveal further details about the disc
structure and AaAb orbit in a manner similar to KH~15D
\cite{1998AJ....116..261K,2006ApJ...644..510W}.

Using the current best-fit visual orbit for AB
\cite{2014AJ....147..123T}, the radial velocity of BaBb should be higher
at the epoch of our ALMA observations (2017.874) than in the last few
decades, because it is now closer to AaAb and moving more slowly towards
Earth. However, the velocity of 5.1~km~s$^{-1}$ at epoch 2017.874 is
lower than 5.73~km~s$^{-1}$ at 1991.96 found for BaBb by ref
\cite{1995ApJ...452..870T}. That is, relative to the system centre of
mass BaBb is moving towards Earth more rapidly now than it was in the
1990s. A possible reason is that the orbit is more eccentric than the
best-fit visual orbit suggests.

To derive an updated orbit, we obtained previous observations of the AB
separation and position angle from the Washington Double Star catalogue
(WDS) \cite{2001AJ....122.3466M}. The system does not appear in Gaia DR2,
presumably because of the multiple nature of the system. These data
extend back to the early 1900s, and all observations before 2009 have no
uncertainties. Based on the scatter from fitting results, we estimated
pre-1950 observations to have position angle uncertainties of 2$^\circ$,
1950-2009 observations to have uncertainties of 1$^\circ$, and all
pre-2009 separations to have uncertainties of 0.1\arcsec. Post-2009
position angle uncertainties were assumed to be 0.5$^\circ$, the
estimated systematic uncertainty \cite{2014AJ....147..123T}, and
separation uncertainties were used as given. Two further observations
are the $7.02 \pm 0.2$~km~s$^{-1}$ difference between the radial
velocity of A and B at epoch 1991.96, and the -0.61~km~s$^{-1}$
difference between the 1991.96 and 2017.874 velocities for B, which
assumes that the radial velocity of B is the same as the systemic
velocity $v_{sys}$ derived from the CO modelling.

\begin{center}
	\begin{figure*}
		\includegraphics[width=1\textwidth]{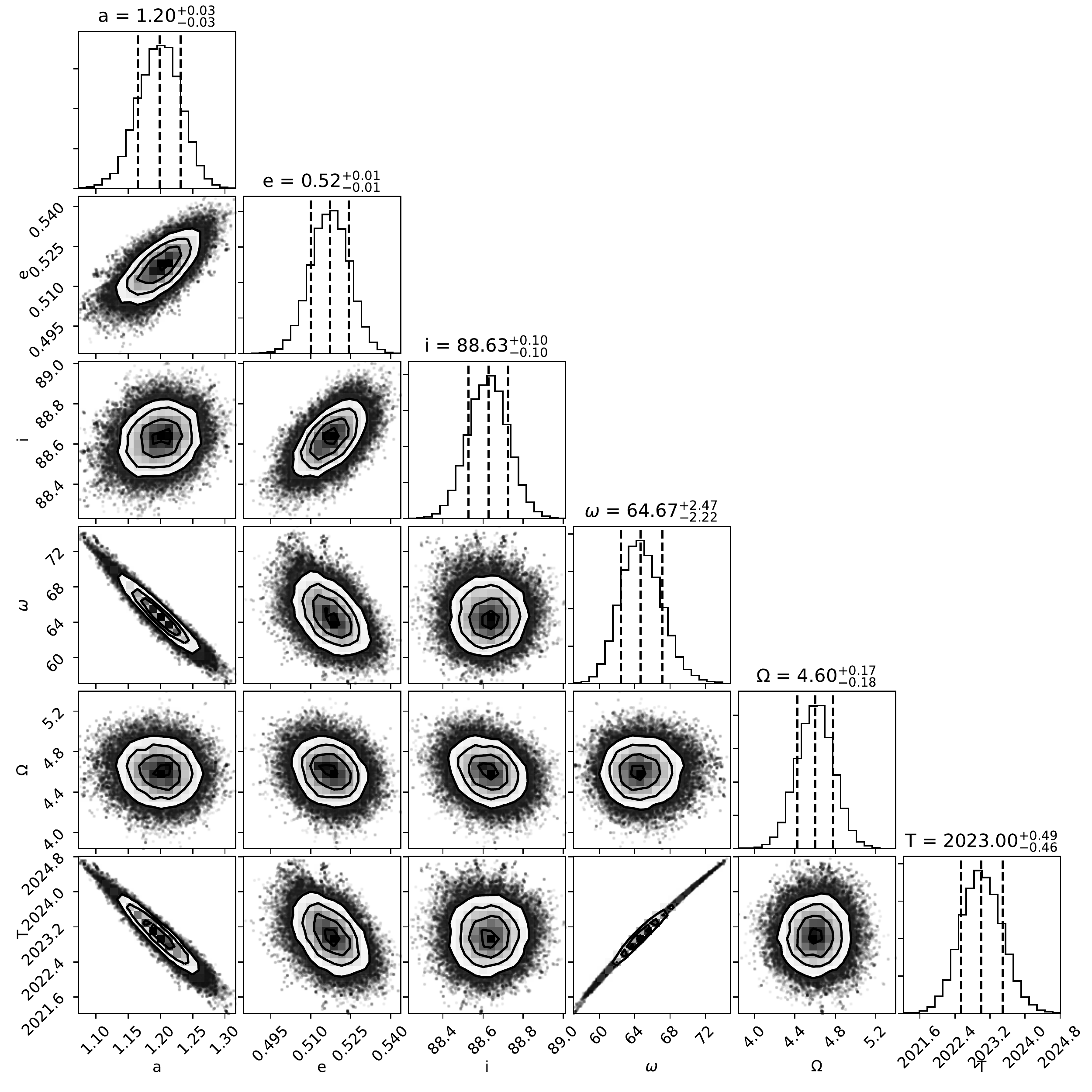}
		\caption{Posterior distributions for the AB orbit. The
                  units of $a$ are seconds of arc, $e$ is dimensionless,
                  $\Omega$ and $\omega$, $i$ are degrees, $T$ is years,
                  and $M_{\rm A}$ is Solar masses.}\label{fig:ab_orb}
	\end{figure*}
\end{center}

To fit an orbit to these data we compute the position angles,
separations, and 1991.96 and 2017.874 radial velocities, from which we
derive a $\chi^2$. As with the visibility modelling described above we
use the python \texttt{emcee} package to find the best-fitting orbits,
and the results are shown in Figure \ref{fig:ab_orb}. The fitted
parameters are the semi-major axis $a$, the eccentricity $e$, the
inclination $i$, the argument of pericentre $\omega$, the ascending node
$\Omega$ (measured anti-clockwise from North), the time of pericentre
passage $T$, and the mass of A $M_{\rm A}$. The mass of A is estimated
as $1.3 \pm 0.15 M_\odot$
\cite{2001ApJ...549..590P,2007ApJ...670.1240A}, and is included as a
parameter in the fitting with a prior reflecting this uncertainty. We
assumed a mass of 1.28$M_\odot$ for B, which is well constrained
\cite{2005ApJ...635..442B}.

Introduction of the additional radial velocity constraint changes the
orbital parameters compared to the most recent published orbit
\cite{2014AJ....147..123T}, which is similar to the most recent orbit
derived by A. Tokovinin (and which has been updated with the correct
ascending node,
\href{http://www.ctio.noao.edu/~atokovin/stars}{http://www.ctio.noao.edu/$\sim$atokovin/stars}).
The main difference relative to these previous orbits are an increased
orbital period (251 vs. 206 years) and eccentricity (0.52 vs. 0.43), and
that uncertainties can now be assigned to each parameter; currently the
AB orbit is graded `5' in the 6th Orbit Catalog \cite{2001AJ....122.3472H},
meaning that the orbital elements `may not even be approximately
correct' (however the proximity to our updated solution suggests that in
hindsight a higher grade could have been assigned). There remain small
systematic offsets in the residual separations and position angles, so
it is possible that the orbit will change somewhat with further
monitoring. We use our best-fit values for the AB orbit in the $n$-body
simulations below.

\begin{center}
	\begin{figure*}
		\includegraphics[width=0.48\textwidth]{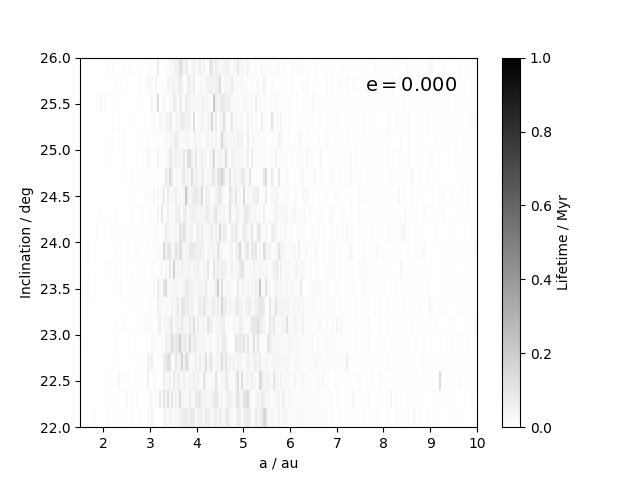}
		\includegraphics[width=0.48\textwidth]{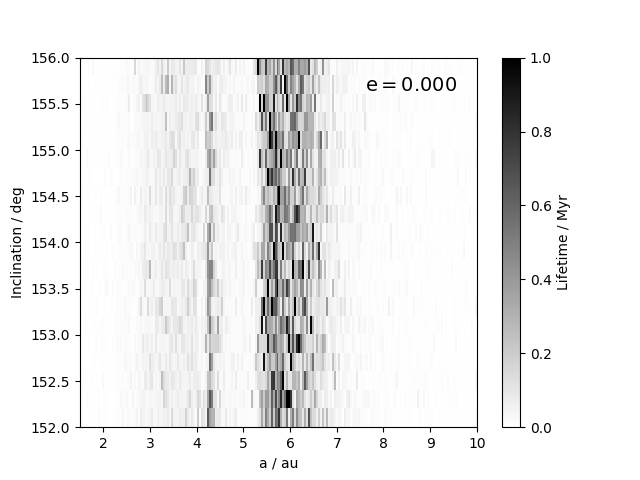}
		\caption{Particle lifetimes for the moderately inclined
                  (left panel) and polar (right panel) disc
                  orientations, shown as a function of semi-major axis
                  and inclination to the sky plane for the best-fit disc
                  orientation. Each pixel shows the lifetime of a single
                  particle with a random initial true anomaly, meaning
                  that no particles in the moderately inclined disc
                  survive for 1Myr. Particles in the polar configuration
                  do survive for up to 1Myr, but only at semi-major axes
                  just outside 4~au and between 5-7au.}\label{fig:nbody}
	\end{figure*}
\end{center}

\subsection{Simulations}\label{sup:sim}

\emph{$n$-body:} We simulated the system as gas-free using the REBOUND
code \cite{2012A&A...537A.128R}, to test where particles could orbit in
the absence of stabilising forces that would arise if the gas mass is
comparable to or greater than the dust mass. We modelled the inner
binary BaBb using the best-fit parameters from ref
\cite{2005ApJ...635..442B}, the outer binary AaAb as a single object,
the AB orbit as derived above, and all other particles as massless. The
disc appears largely circular based on the continuum modelling, so we
initialized particles on circular orbits in one of the two possible disc
orientations at a range of true anomalies, and ran simulations for
1Myr. We used the `whfast' integrator \cite{2015MNRAS.452..376R} with a time
step of 1/20th of the BaBb orbital period. Particles with distances more
than 2000~au from the system centre of mass are removed and their
removal time recorded; while the integration method does not allow
collisions nor compute close encounters accurately, these events soon
lead to particle ejection anyway.

The results are shown in Figure \ref{fig:nbody}, and show that in the
gas-free case test particles do not survive for 1Myr in the moderately
inclined disc configuration. Particles in the polar configuration do
survive for up to 1Myr, but only at semi-major axes between 5 and 7~au
and a very narrow band near 4au, and neither region is consistent with
the extent seen in the continuum with ALMA. In both configurations
particles beyond 7~au are typically removed in 5000~years; comparing
this time to the 250~year period of the AB binary suggests that
short-term interactions are the cause (i.e. the removal time is tens of
AB orbits, not hundreds or thousands which would suggest long-term
secular effects), and therefore that 7~au is the approximate outer disc
truncation radius imposed by A.

These results therefore indicate that gas-free dynamics do not apply to
the dust observed between 2.5 to 4.6~au around HD~98800BaBb, and that
the dust is almost certainly strongly influenced and stabilized by
gas. The disc is therefore probably gas-rich, and given the detection of
hydrogen gas emission \cite{2012ApJ...744..121Y}, likely still in the
primordial protoplanetary disc phase.
 
\begin{center}
	\begin{figure*}
		\includegraphics[width=0.48\textwidth]{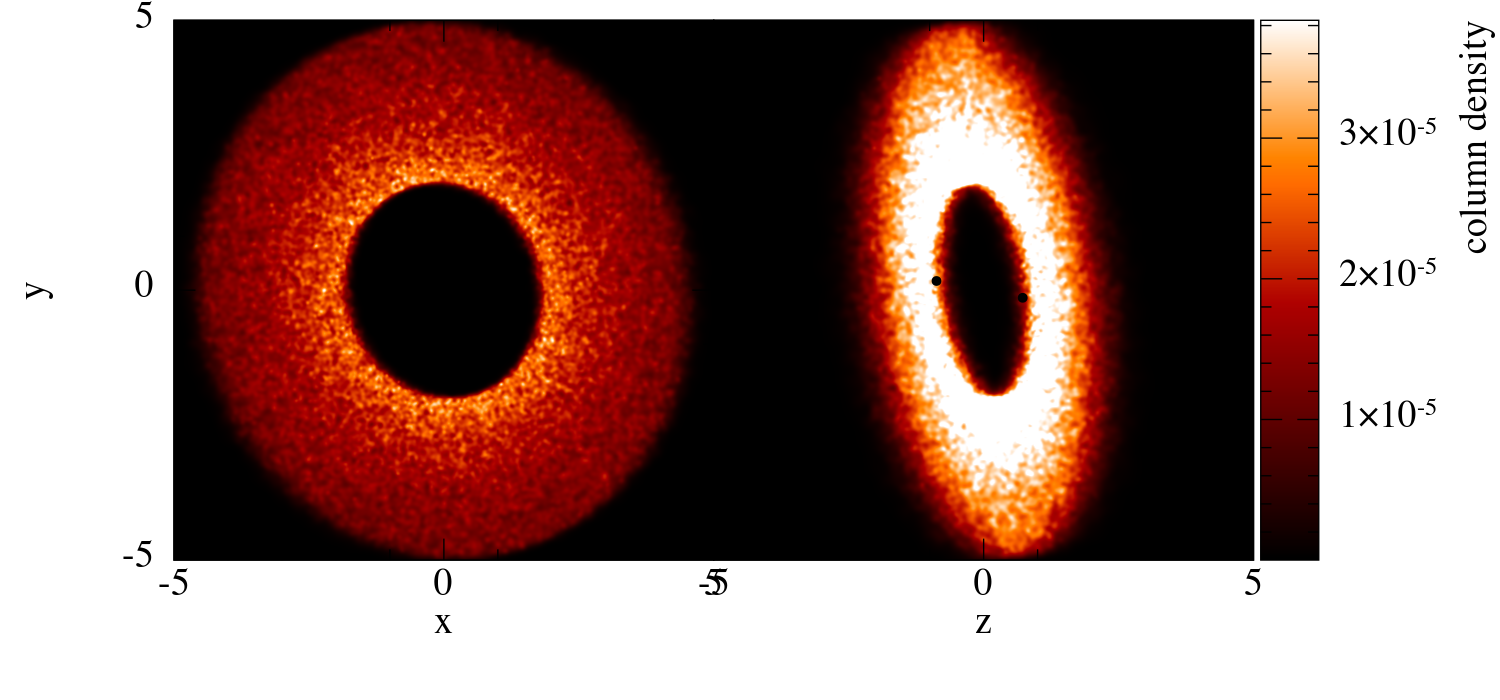}
		\includegraphics[width=0.48\textwidth]{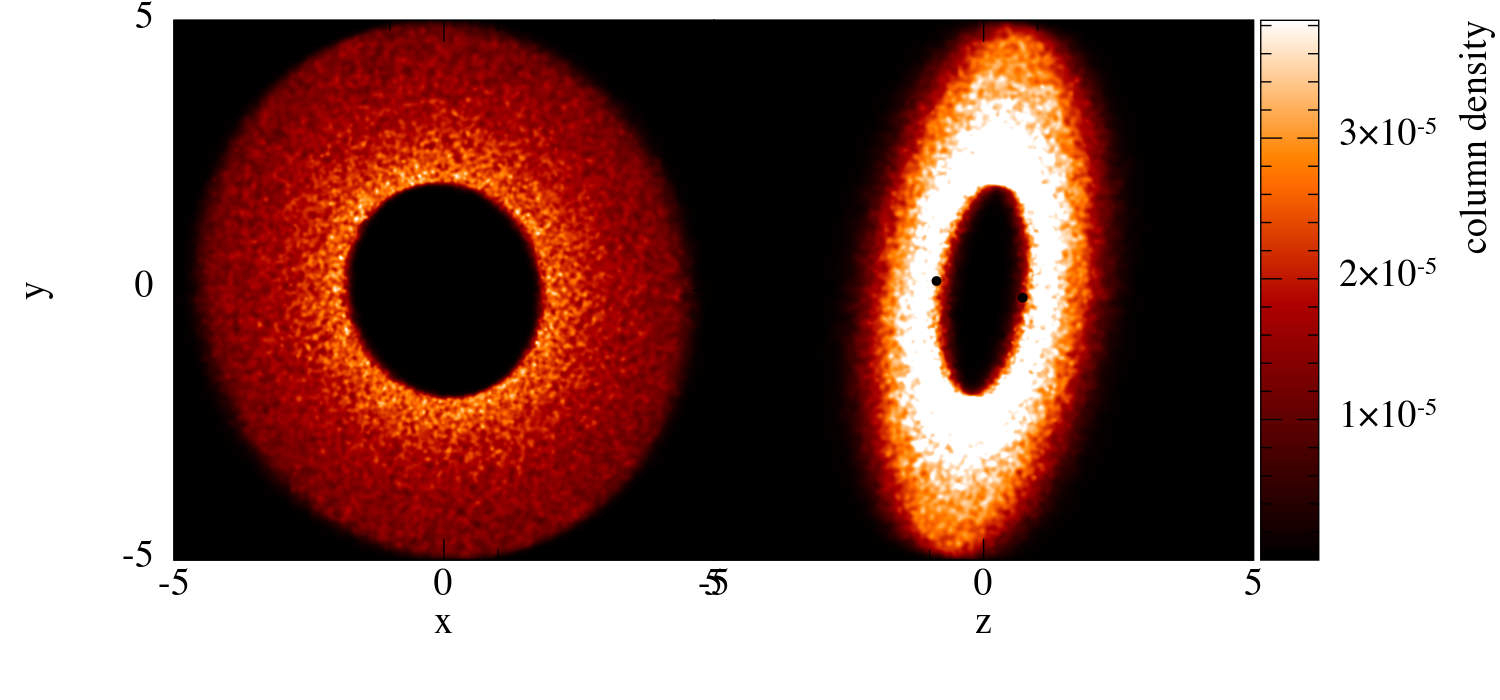}\\
		\includegraphics[width=0.48\textwidth]{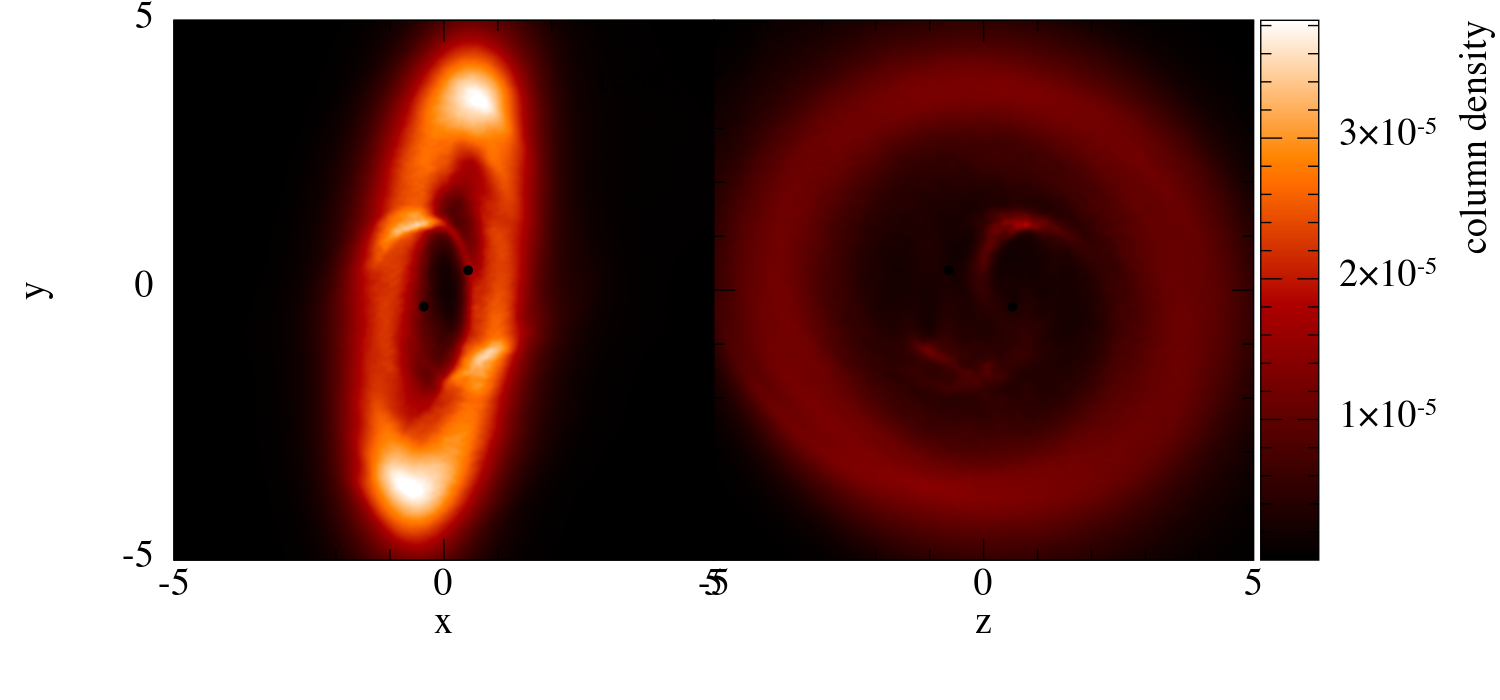}
		\includegraphics[width=0.48\textwidth]{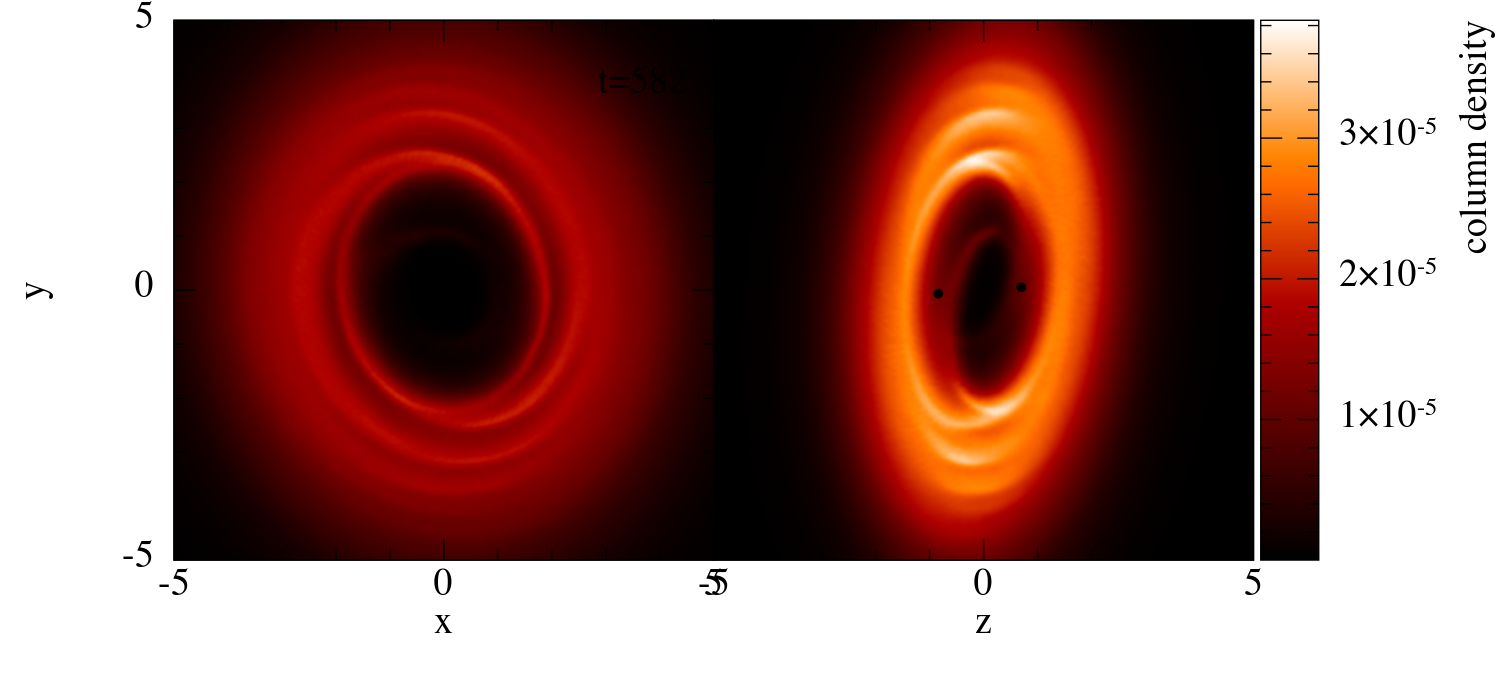}\\
		\includegraphics[width=0.48\textwidth]{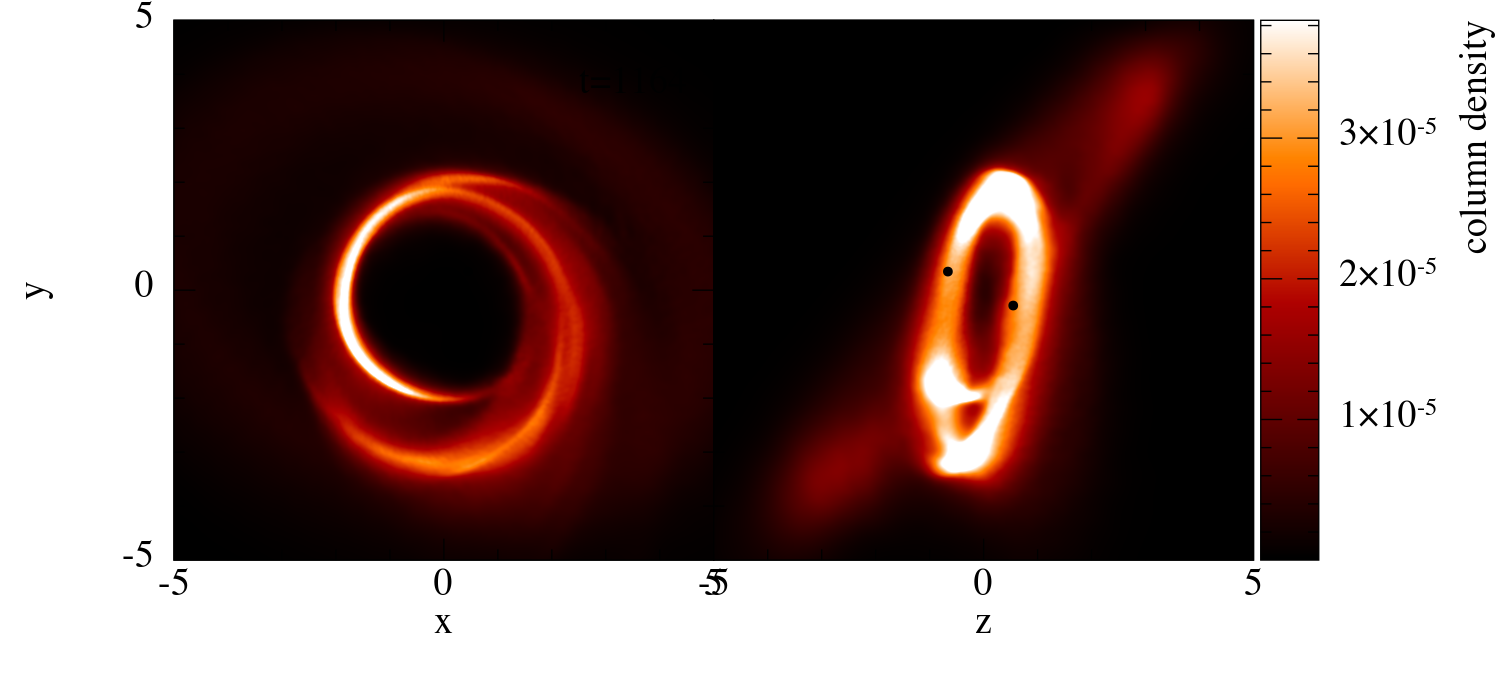}
		\includegraphics[width=0.48\textwidth]{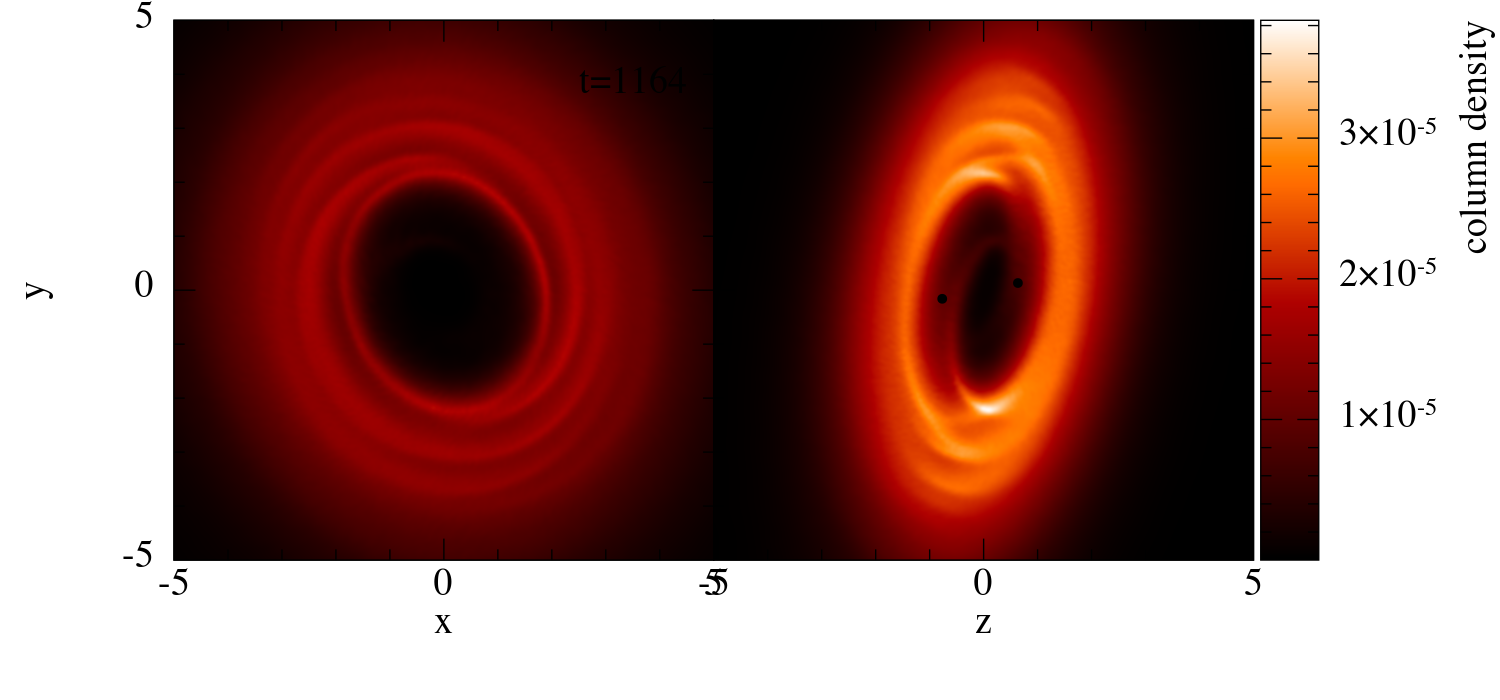}\\
		\includegraphics[width=0.48\textwidth]{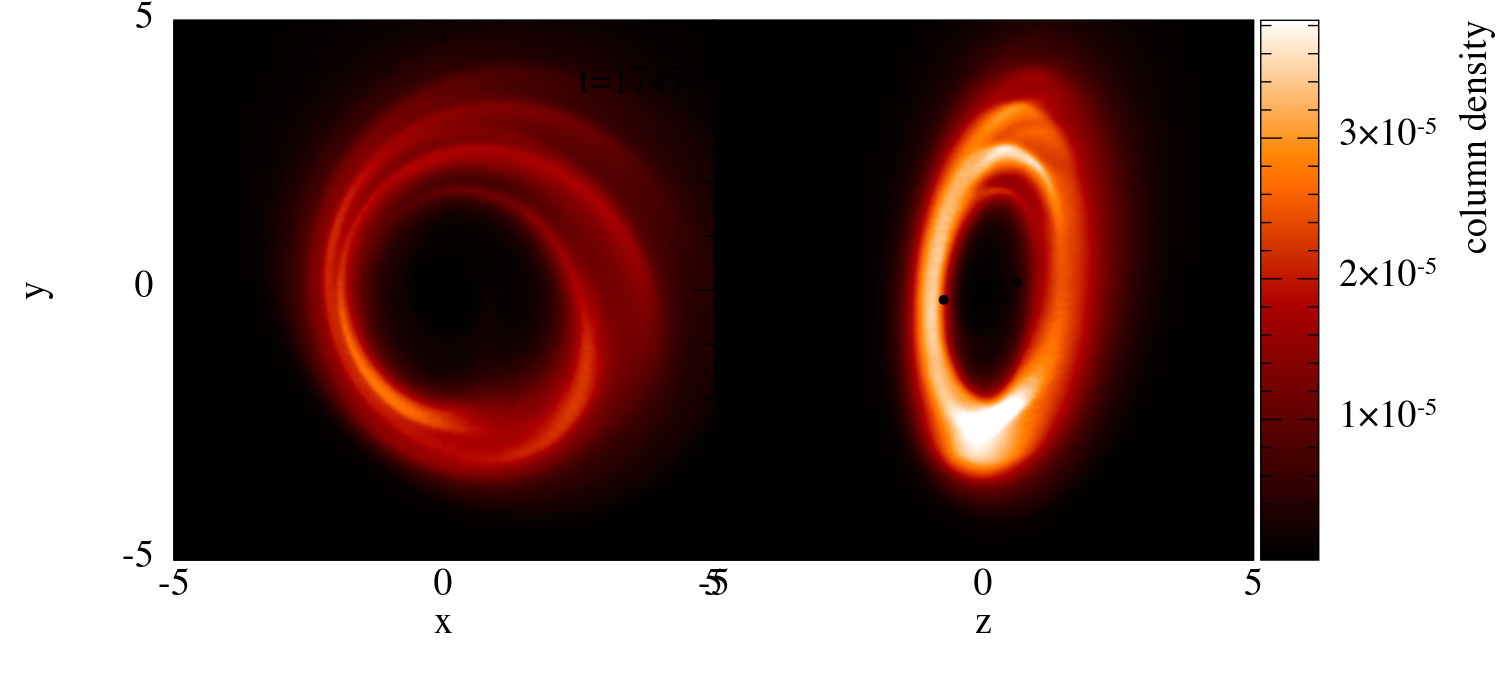}
		\includegraphics[width=0.48\textwidth]{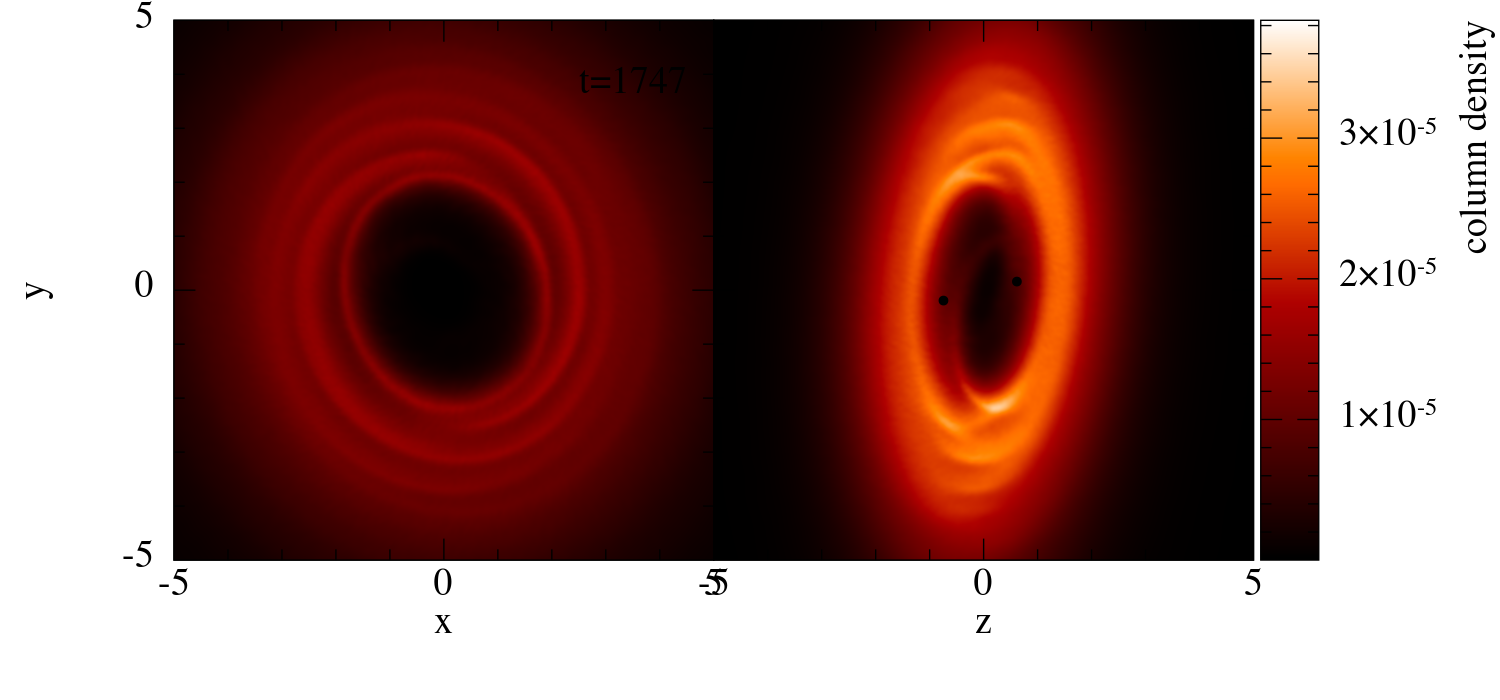}\\
		\includegraphics[width=0.48\textwidth]{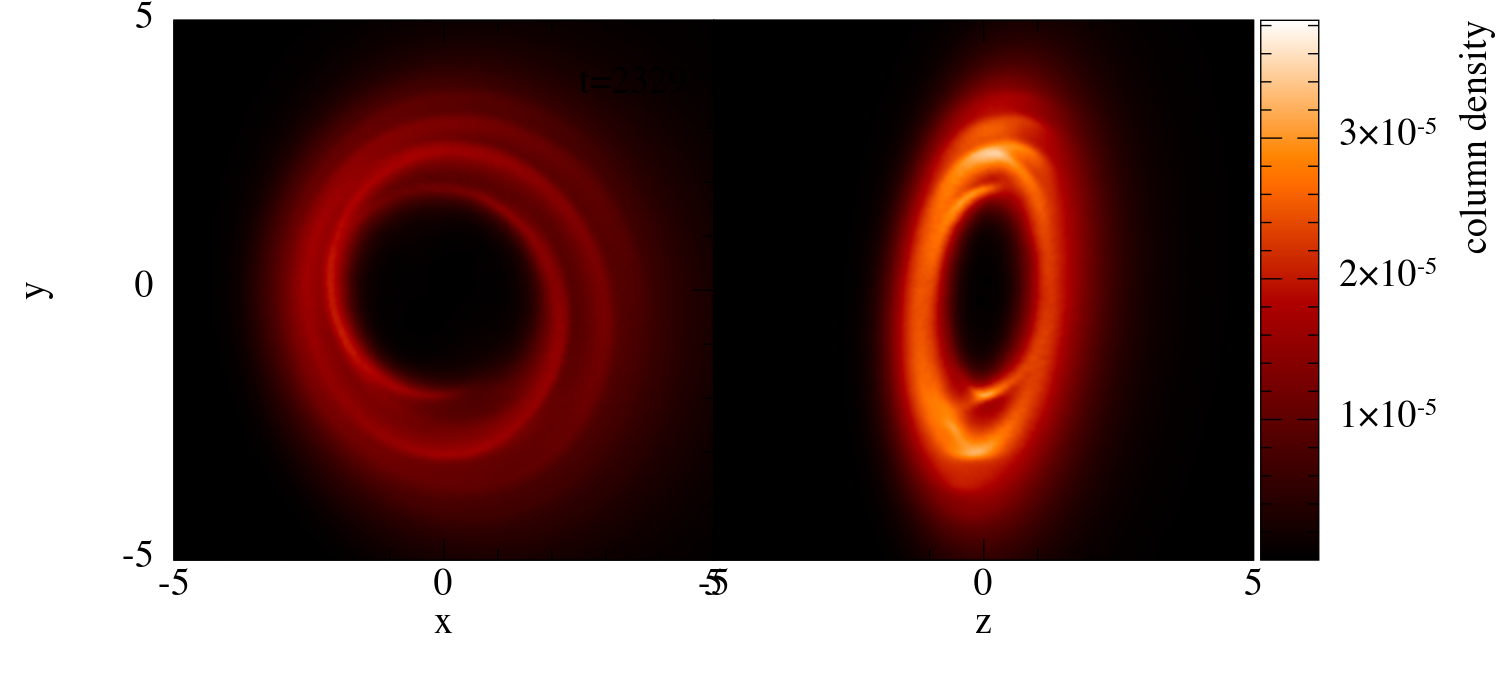}
		\includegraphics[width=0.48\textwidth]{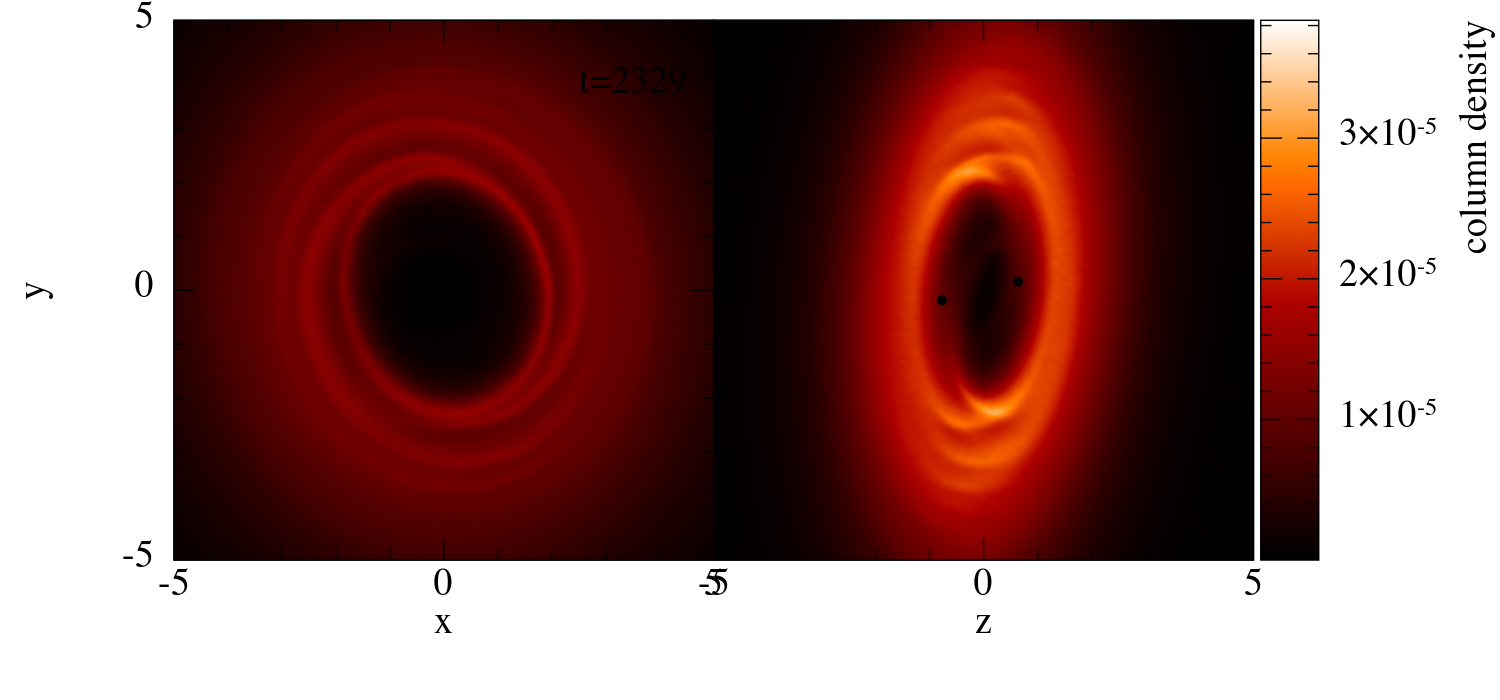}\\
                \vspace{-2cm}
		\caption{SPH simulations of the disc in the moderately
                  misaligned (left panels) and polar (right panels)
                  configurations, where each panel is separated by 100
                  binary orbits (about 86 years) in time, from top to
                  bottom. The colour scale shows gas column density. In
                  each panel, the x/y axes correspond to the sky plane
                  in au, and the z to the line of sight (with +ve z
                  towards Earth). The moderately misaligned disc
                  precesses significantly over the duration of the
                  simulation, eventually reaching a similar orientation
                  to the disc in the polar
                  configuration.}\label{fig:sph}
	\end{figure*}
\end{center}

\emph{Smoothed Particle Hydrodynamics:} We used Phantom
\cite{2018PASA...35...31P} to simulate the response of the two possible
disc configurations to perturbations from the inner binary, assuming a
gas-rich protoplanetary disc. We did not include the outer binary, as
the disc was found to change orientation on a shorter timescale
($\sim$100 years) than the period of the outer binary ($\approx$250
years). This timescale difference of course does not mean that A has no
effect on the disc around B, since test particles in the polar
configuration are truncated at around 7~au, which is near the outer
extent of the detected CO.

For the disc, the simulations used the same SPH parameters as ref
\cite{2017ApJ...835L..28M} and 300,000 particles. The results of
simulations of the two configurations are shown in Figure \ref{fig:sph},
where the left set of panels shows the moderately misaligned case, and
the right panels the polar case (the images were created using
\texttt{splash} \cite{2007PASA...24..159P}). In each pair of images, the
left subpanels show the disc density projected onto the sky plane, and
the right panel a side view. The time evolution (moving down the panels)
shows that the polar configuration does not significantly change
orientation, while the moderately misaligned case does. In fact, the
moderately misaligned disc eventually reaches the polar configuration,
though has been significantly perturbed and disrupted in the
process. Thus, we conclude that the polar disc configuration is by far
the best interpretation of the data, as the lifetime of the disc in the
moderately misaligned case is very short relative to the stellar age of
ten million years. Because the time taken to re-orient the disc is short
relative to the AB orbital period, and the effect of A where the disc is
observed is minimal, including A in the simulations would not change
this conclusion.

Some spiral structure has been induced in the disc in the polar case; if
the HD~98800 disc has similar structure these perturbations may be the
cause of the asymmetry that requires the continuum model to have
non-zero $e_{in}$.

%\subsection{SPHERE ZIMPOL imaging}\label{sup:sphere}

%\textbf{This will go in if Julien can make some images!}

% In the course of this investigation we also obtained an image of the
% HD~98800 system with the Spectro-Polarimetric High-contrast Exoplanet
% REsearch facility (SPHERE) \cite{2008SPIE.7014E..18B} using the Zurich
% Imaging Polarimeter (ZIMPOL) \cite{2008SPIE.7014E..3FT}.

\begin{center}
	\begin{figure*}
		\includegraphics[width=1\textwidth]{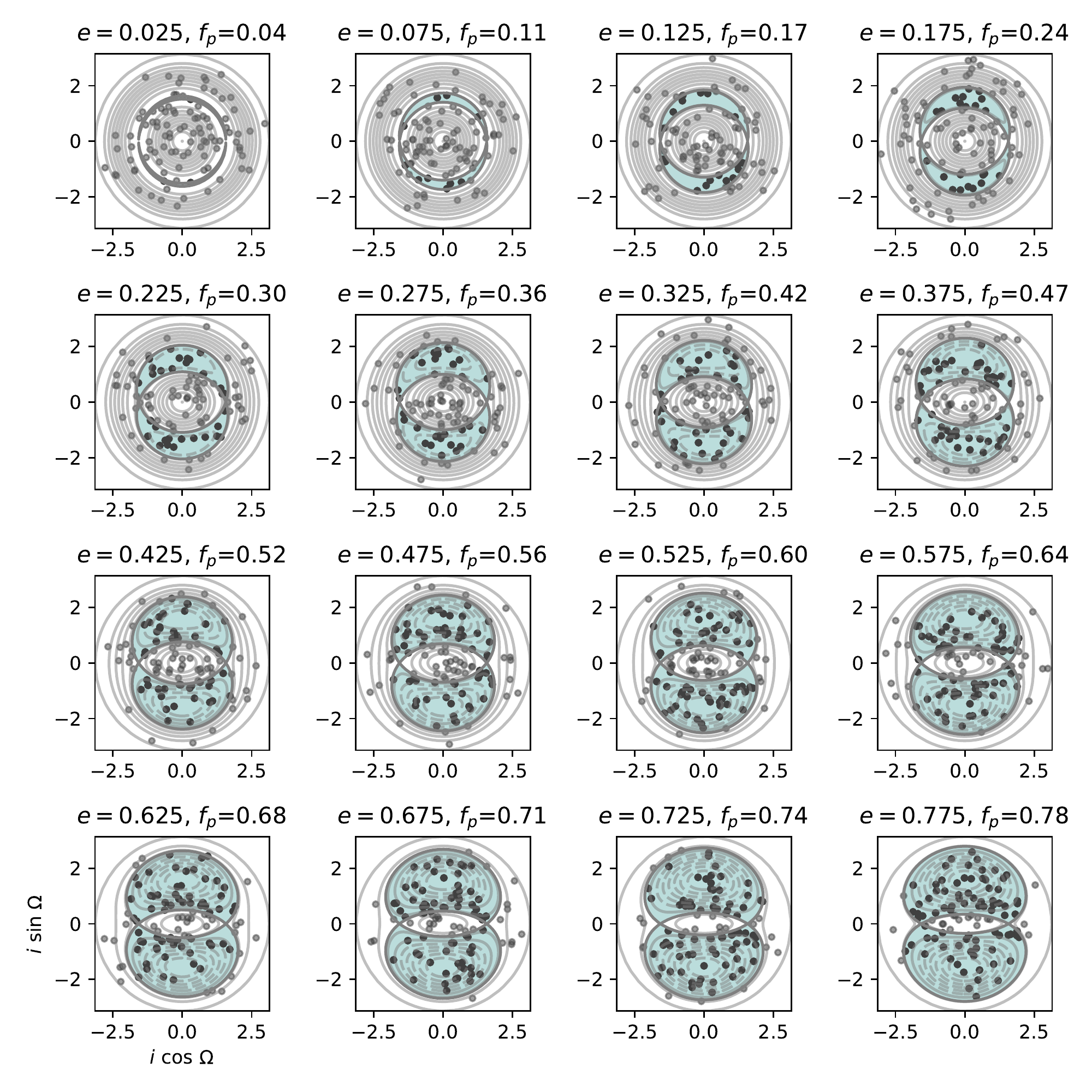}
		\caption{Model of the fraction of uniformly oriented
                  circumbinary discs that evolve to polar orbits, as a
                  function of eccentricity. Each panel shows contours of
                  the constant of motion $c$ for circumbinary orbits
                  uniformly spaced between $-4e^2$ and 1. Polar orbits
                  lie in the grey regions (dashed contours), and
                  coplanar orbits lie in the white regions (solid
                  contours). Dots show a subsample of 100 random
                  orientations, with coplanar configurations shown by
                  grey dots, and polar configurations by black dots. The
                  title of each sub-panel shows the binary eccentricity
                  and the fraction of polar
                  orbits.}\label{fig:polar_frac}
	\end{figure*}
\end{center}

\subsection{Population estimate of disc orientations}\label{sup:pop}

To gauge whether circumbinary discs in the polar configuration might be
common, we make a simple population estimate. The assumptions are that
i) the discs do not have sufficient mass to re-orient the binary
(i.e. test-particle dynamics apply), ii) discs initially have uniformly
distributed orientations with respect to the binary, iii) the binary
eccentricity distribution is uniform between 0 and 0.8
\cite{2010ApJS..190....1R}. The most uncertain of these is ii); if
circumbinary discs are more often initially coplanar with their binaries
the estimated fraction would be lower than we find below.

The evolution of a low-mass circumbinary disc can be visualized in the
$i \cos{\Omega}$, $i \sin{\Omega}$ plane with plots like those shown in
Figure \ref{fig:polar_frac}, where the constant of motion for a given
disc is \cite{2010MNRAS.401.1189F}
\begin{equation}
  c = \cos^2i - e^2 \sin^2i \left( 5 \sin^2\Omega - 1 \right) \, .
\end{equation}
The dividing line (`separatrix') between orbits that are coplanar
(i.e. those that precess about the binary orbital plane) and polar
(i.e. those that precess about a plane perpendicular to the binary
pericentre direction) is given by $c = e^2$. We assume here that the
region of parameter space in which a disc initially starts determines
the outcome; dissipation causes the disc plane to relax to coplanar
(0,0), or polar (0,$\pm \pi/2$), and discs do not cross the separatrix
\cite{2017ApJ...835L..28M,2018MNRAS.473..603Z}.

To estimate the fraction of discs on coplanar and polar orbits for a
given eccentricity, we populate the $i \cos{\Omega}$, $i \sin{\Omega}$
plane with $N=10^6$ orbits. Orbits are uniformly distributed by
generating two random numbers $u$ and $v$ between 0 and 1 for each
orbit, from which we obtain $\Omega = 2 \pi u$ and
$i = \cos^{-1}(2v-1)$. We then calculate $c$ for each of the $N$ orbits,
and find the fraction that lie on polar trajectories (i.e. the fraction
in the grey region in Figure \ref{fig:polar_frac}).

We then repeat the calculation at a series of 16 binary eccentricities
between 0.0 and 0.8, as shown in Figure \ref{fig:polar_frac}. The
overall fraction of discs that should evolve to polar configurations
given our assumptions is the average value across this range of
eccentricities, and found to be 46\%. The systems that evolve to polar
configurations are heavily weighted towards high eccentricities, as the
fraction of parameter space where this happens is much larger. As noted
in the main text, the eccentricities of the binaries that host
circumbinary planets are typically small, and higher eccentricity
binaries are the most promising focus for misaligned circumbinary planet
searches.

\subsection{Data and code availability}\label{sup:avail}

The ALMA data used in this study are available in the ALMA Science
Archive at
\href{http://almascience.eso.org/aq/}{http://almascience.eso.org/aq/}.

The post-processing, modelling, and other scripts used in this study are
available on github at
\href{https://github.com/drgmk}{https://github.com/drgmk}.

\end{methods}

%\bibliographystyle{naturemag}
%\bibliography{../../ref} % if your bibtex file is called example.bib

\subsection{Acknowledgements}

GMK is supported by the Royal Society as a Royal Society University
Research Fellow. L.M. acknowledges support from the Smithsonian
Institution as a Submillimeter Array (SMA) Fellow. OP is supported by
the Royal Society Dorothy Hodgkin Fellowship. We thank A. Ribas for
sharing the VLA image of HD~98800.

This paper makes use of the following ALMA data:
ADS/JAO.ALMA\#2017.1.00350.S. ALMA is a partnership of ESO (representing
its member states), NSF (USA) and NINS (Japan), together with NRC
(Canada), MOST and ASIAA (Taiwan), and KASI (Republic of Korea), in
cooperation with the Republic of Chile. The Joint ALMA Observatory is
operated by ESO, AUI/NRAO and NAOJ.

\subsection{Author Contributions}

GMK conceived the project, analysed the data, carried out the modelling,
and wrote the manuscript. LM contributed gas calculations and provided
advice on self-calibration. BMY set up and ran the $n$-body
simulations. DP provided advice on running the SPH simulations. All
co-authors provided input on the manuscript.

\subsection{Author Information}

The authors declare that they have no competing
interests. Correspondence and requests for materials should be addressed
to g.kennedy@warwick.ac.uk.

\end{document}